\documentclass[review,journal]{IEEEtran}
\usepackage{amsmath,amsfonts}
\usepackage{algorithmic}
\usepackage{algorithm}
\usepackage{endnotes}
\usepackage{array}
\usepackage{adjustbox}
\usepackage{textcomp}
\usepackage{stfloats}
\usepackage{subcaption}
\usepackage{setspace}
\usepackage{url}
\usepackage{verbatim}
\usepackage{graphicx}
\usepackage[hidelinks]{hyperref}
\hypersetup{
    colorlinks=true,
    linkcolor=blue,
    citecolor=blue,
    filecolor=magenta,      
    urlcolor=cyan,
}
\usepackage{cite}
\usepackage{booktabs}
\usepackage{array}
\usepackage{graphicx}
\usepackage{booktabs} %
\usepackage[table,xcdraw]{xcolor} %
\usepackage[table]{xcolor} %
\usepackage{multirow} %
\usepackage{geometry}
\usepackage{tabu}

\usepackage[table, xcdraw]{xcolor} %
\usepackage{multirow}
\usepackage{array}
\usepackage{colortbl}
\usepackage{caption}

\usepackage{tikz}
\usetikzlibrary{trees}

\definecolor{DarkYellow}{HTML}{DEA601}
\definecolor{DarkOrange}{HTML}{ED7D31}
\definecolor{DarkBlue}{HTML}{4472C4}
\usepackage{xspace}

\newcommand{\taskWhy}{\textit{\textbf{\textcolor{DarkOrange}{Why?}}}\xspace}
\newcommand{\taskHow}{\textit{\textbf{\textcolor{DarkBlue}{How?}}}\xspace}
\newcommand{\taskWhat}{\textit{\textbf{\textcolor{DarkYellow}{What?}}}\xspace}

\newcommand{\websiteURL}{\url{https://shape-vis.github.io/annotation_star/}}

\definecolor{indvHigh}{HTML}{9BB2E0}
\definecolor{indvMed}{HTML}{C8D6ED}
\definecolor{indvLow}{HTML}{ECF2F9}

\definecolor{grpHigh}{HTML}{B6E09B}
\definecolor{grpMed}{HTML}{D6EDC7}
\definecolor{grpLow}{HTML}{F2F9ED}

\definecolor{ensTaskHigh}{HTML}{B86029}
\definecolor{ensTaskMed}{HTML}{EAB38B}
\definecolor{ensTaskMedLow}{HTML}{EFCAB0}
\definecolor{ensTaskLow}{HTML}{F8E4D8}

\definecolor{ensTotHigh}{HTML}{B7912F}
\definecolor{ensTotMed}{HTML}{F5C242}
\definecolor{ensTotMedLow}{HTML}{F8DA78}
\definecolor{ensTotLow}{HTML}{FDF3D0}

\definecolor{dsHigh}{HTML}{2F5597}
\definecolor{dsMed}{HTML}{8FAADC}
\definecolor{dsLow}{HTML}{B4C7E7}

\begin{document}

\title{A Survey on Annotations in Information Visualization: Empirical Insights, Applications, and Challenges}

\author{Md Dilshadur Rahman, Bhavana Doppalapudi, Ghulam Jilani Quadri, Paul Rosen
\thanks{Md Dilshadur Rahman and Paul Rosen are with the University of Utah. E-mail: \{dilshadur,prosen\}@sci.utah.edu}%
\thanks{Bhavana Doppalapudi is with the University of South Florida. E-mail: bhavanadoppalapudi@gmail.com}%
\thanks{Ghulam Jilani Quadri is with the University of Oklahoma. E-mail: quadri@ou.edu}%
\thanks{Manuscript received April 19, 2021; revised August 16, 2021.}}

\markboth{Journal of \LaTeX\ Class Files,~Vol.~14, No.~8, August~2021}%
{Shell \MakeLowercase{\textit{et al.}}: A Sample Article Using IEEEtran.cls for IEEE Journals}

\maketitle

\begin{abstract}
We present a comprehensive survey on the use of annotations in information visualizations, highlighting their crucial role in improving audience understanding and engagement with visual data. Our investigation encompasses empirical studies on annotations, showcasing their impact on user engagement, interaction, comprehension, and memorability across various contexts. We also study the existing tools and techniques for creating annotations and their diverse applications, enhancing the understanding of both practical and theoretical aspects of annotations in data visualization. Additionally, we identify existing research gaps and propose potential future research directions, making our survey a valuable resource for researchers, visualization designers, and practitioners by providing a thorough understanding of the application of annotations in visualization. An interactive web resource detailing the surveyed papers is available at \websiteURL.
\end{abstract}

\begin{IEEEkeywords}
Survey, Visualization, Annotation.
\end{IEEEkeywords}

\setstretch{0.955}

\section{Introduction}
\label{sec.intro}

Annotations are a vital component of visualizations, serving as guides, providing additional context, emphasizing critical key data features, and enhancing data comprehension. Although there remains no consensus definition for what an annotation is, they are, roughly speaking, graphical or textual elements added to a visualization that serves as additional attributes for associated data items~\cite{munzner2014visualization}. Annotations help structure viewers' mental models as they use visualizations to understand the insights behind the data~\cite{cedilnik2000procedural}. Furthermore, annotating charts have been recognized as a critical task facilitating visual data analysis~\cite{heer2012interactive, zhao2016annotation}, crucial for externalizing and exploring data~\cite{kang2014characterizing, sevastjanova2021visinreport, mahyar2012note, shrinivasan2009connecting, shrinivasan2008supporting, kim2019inking, choe2015characterizing, heer2012interactive, lin2022data, romat2019activeink, williams2023data}, aiding collaborative sensemaking~\cite{sanderson1994exploratory, chen2011supporting, chen2010click2annotate, mahyar2012note, mahyar2014supporting, isenberg2006interactive, robinson2008collaborative, kong2009perceptual, ellis2004collaborative}, and contributing to narrative storytelling~\cite{segel2010narrative, lee2015more, kosara2013storytelling, hullman2011visualization, satyanarayan2014authoring, ge2020canis, rahman2023exploring}.

\begin{figure}[!b]
  \centering
  \includegraphics[width=0.975\linewidth]{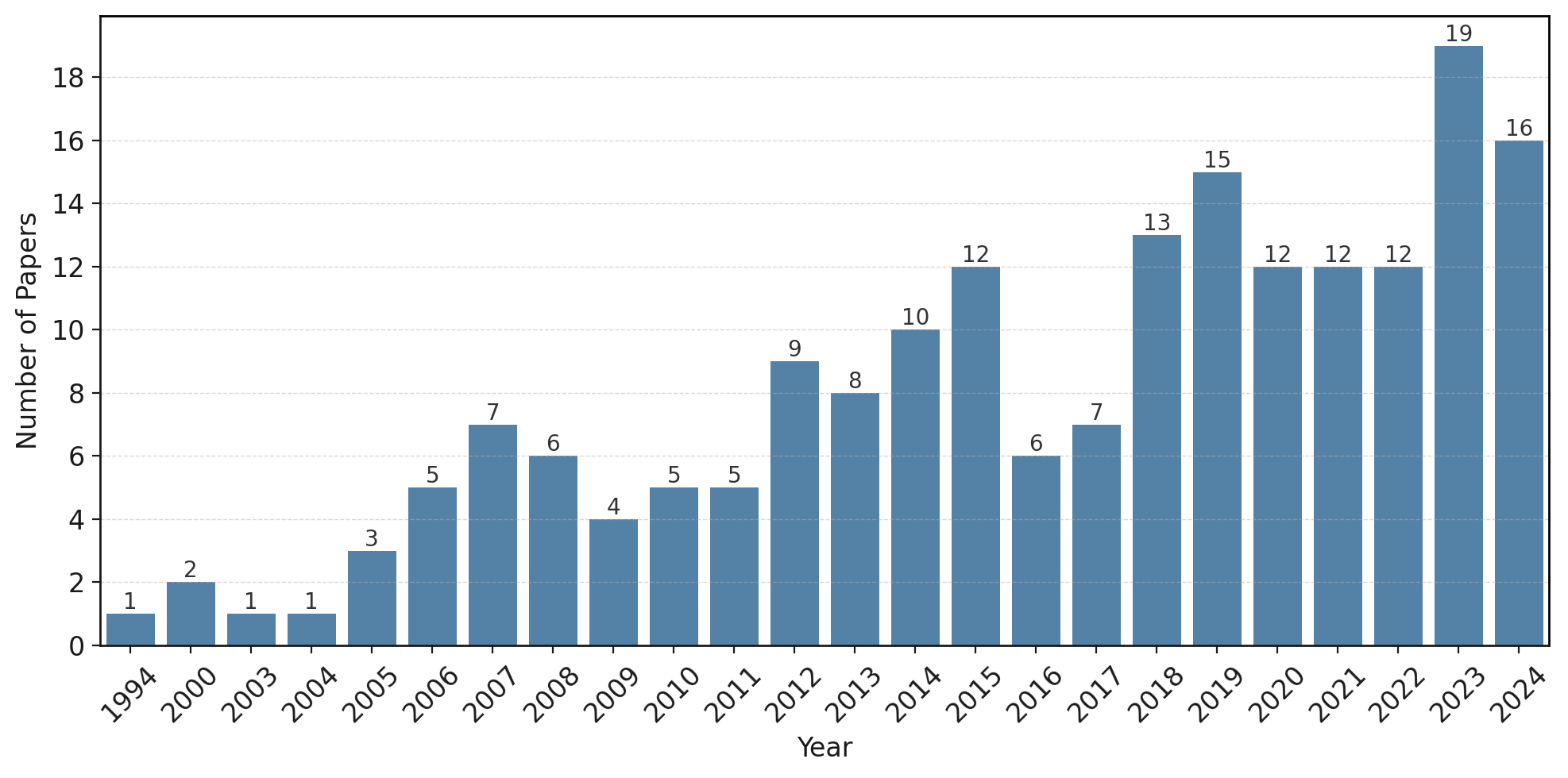} 
  \caption{The chart of publications discussing the importance, application, and tools facilitating annotations in visualizations by year illustrates a noteworthy increase.
  }
  \label{fig:papers_by_year}
\end{figure}

\begin{figure}[ht]
    \centering
    \includegraphics[width=0.8\linewidth]{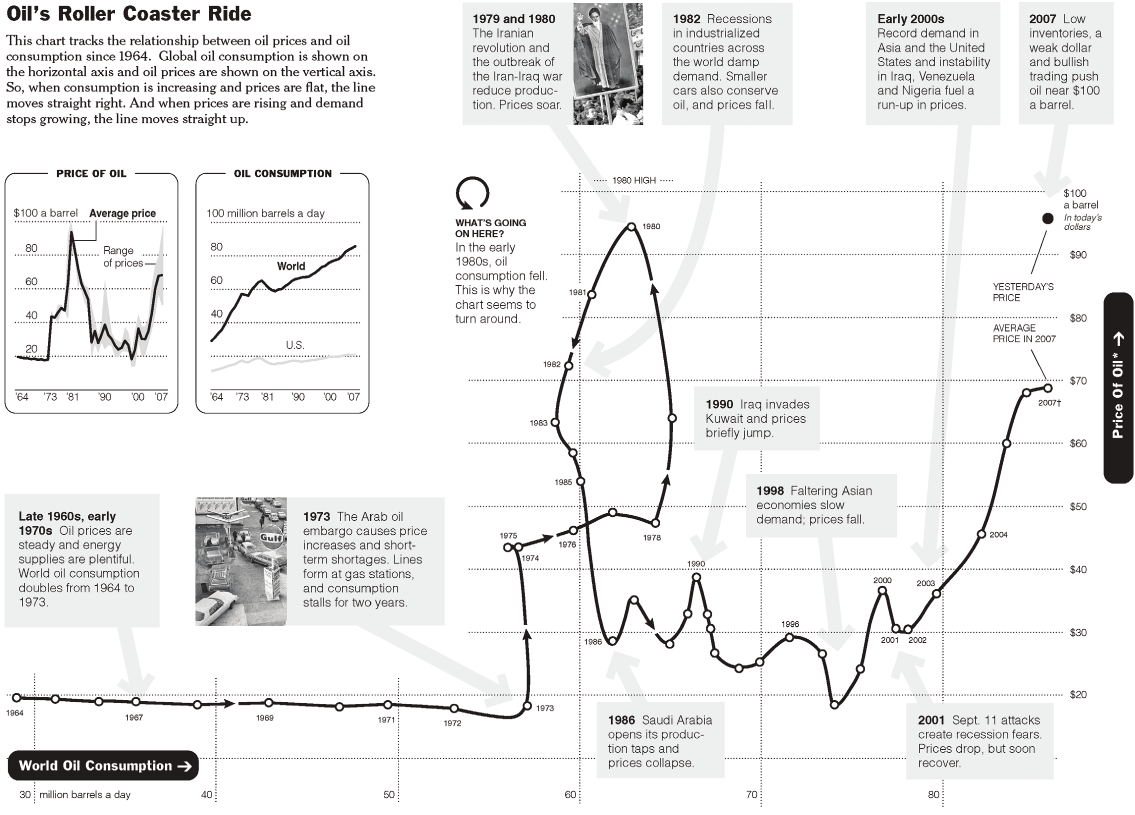}
    \caption{Connected scatterplot from the New York Times showing the correlation between oil prices and consumption (1964–2007)~\cite{10.1111/j.1740-9713.2012.00605.x}. Key events are annotated with text, rectangles, and arrows.}
    \label{fig:nyt-conn-scatter}
\end{figure}

Over the past two decades, many commercial interactive data visualization tools, such as Tableau, Microsoft PowerBI, and Google Data Studio, have emerged, supporting the annotation of visualizations and dashboards with textual and graphical elements. Notably, significant research in annotation within the visualization community has been inspired by the work of practitioners. For instance, early work by Segel et al.~\cite{segel2010narrative} explored the role of annotations in narrative visualizations, focusing on how data journalists use annotations for storytelling. Heavily annotated charts from the New York Times graphics team, such as the connected scatterplots by Amanda Cox, Hannah Fairfield, and others~\cite{cox_nguyen_oil_prices_2008, fairfield_driving_reverse_2010} (see~\autoref{fig:nyt-conn-scatter}), have inspired research on the utility of annotations for better comprehension~\cite{haroz2015connected}, and Ren et al.'s work is inspired by the New York Times' interactive chart-making tool, Mr. Chartmaker~\cite{mrchartmaker_2015}. Furthermore, publication trends over the past twenty years indicate a growing interest in annotations within the visualization community (see~\autoref{fig:papers_by_year}). The number of research articles on annotations, including tools for authoring annotations~\cite{ren2017chartaccent, chen2010click2annotate}, techniques for various annotation approaches~\cite{cedilnik2000procedural, badam2022integrating}, and empirical studies on different aspects of annotating visualizations~\cite{stokes2022striking, borkin2015beyond}, has significantly increased. This trend underscores the rising recognition of annotations' importance in improving visual data interpretation and interaction.

Despite the widespread use and recognized importance of annotations in visualizations, there is a clear gap: the lack of a comprehensive survey on this topic. To address this, we present a systematic literature survey on annotations in information visualizations, examining developments over the past two decades. Our survey reviews existing design spaces of annotations, annotation types, and empirical studies that explore their application in visualizations. We also synthesize annotation techniques and features of tools that support annotation authoring across different contexts. Additionally, we discuss insights, takeaways, and implications for developing better annotation tools. We identify challenges and highlight areas needing further investigation. Our survey will provide researchers with a comprehensive understanding of the current tools, techniques, and practices in annotation, providing a foundation to identify new research opportunities. For practitioners, the insights offer a clear understanding of available annotation methods and tools, enabling them to improve their visualization design processes and outputs.

\section{Methodology}
\label{sec.method}

\begin{figure}[!b]
  \centering
  \includegraphics[width=0.85\linewidth]{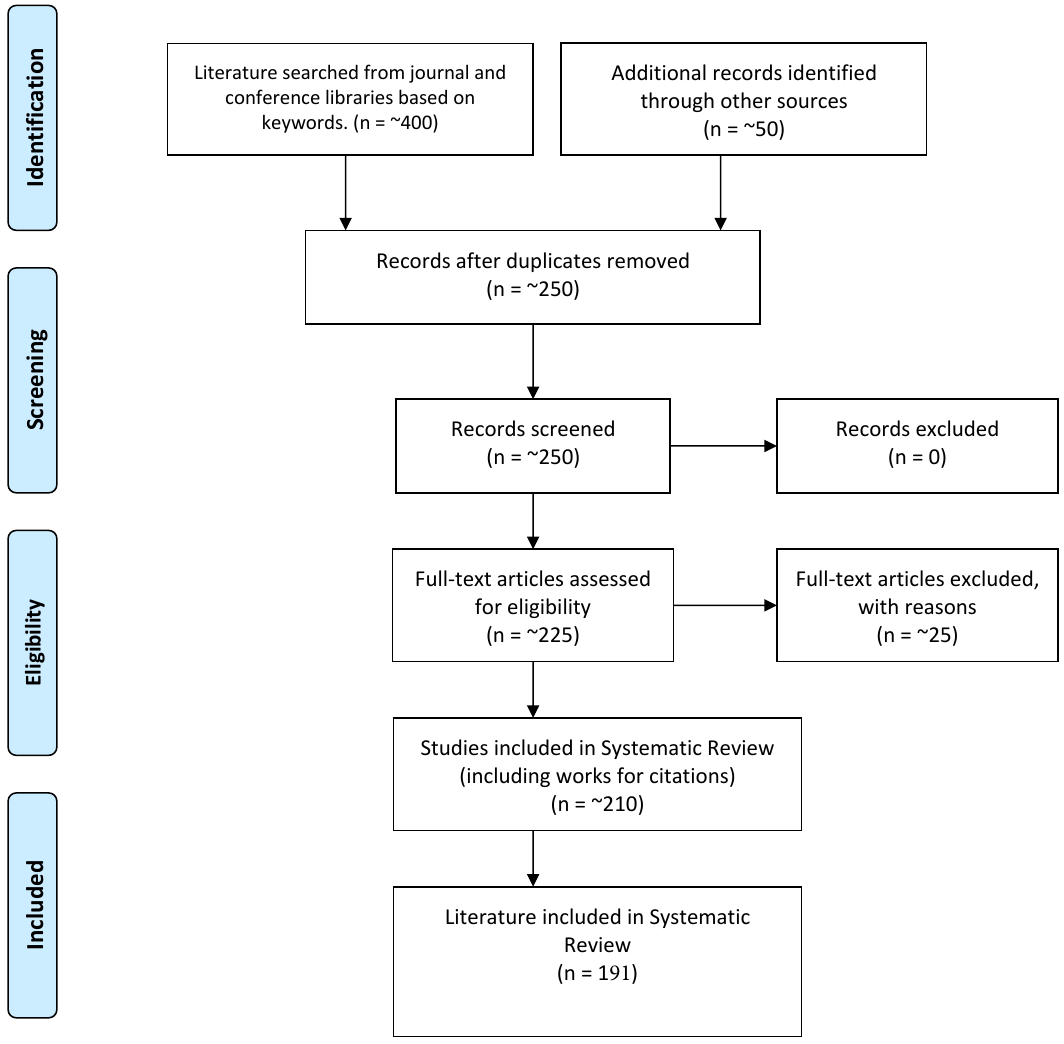} 
  \caption{Prisma Framework for literature selection.}
  \label{fig:prisma}
\end{figure}

We conducted a systematic literature search conducted independently by two co-authors identifying papers using the following keywords: \textit{annotate}, \textit{annotated charts}, \textit{annotation}, and \textit{annotations in visualizations} from academic databases that included ACM, IEEE, EG/CGF, and Sage (Information Visualization Journal). We also used \url{vispubs.com}~\cite{langevispubs} to find papers published in relevant venues. The initial search yielded approximately 450 unique papers. Subsequently, one co-author reviewed each paper to assess its relevance to our survey based on the inclusion and exclusion criteria discussed next, applying the PRISMA framework (see~\autoref{fig:prisma}). This process resulted in a corpus of 191 papers sourced mainly from leading visualization journals, conferences, and symposia (see~\autoref{tab:source}).

\vspace{2pt}
\noindent
The inclusion criteria of our survey include:

\begin{itemize}
    \item Studies discussing the definition and design spaces of annotations in visualizations and studies providing foundational knowledge on annotation types, generation methods, targets, and their importance in visualizations (e.g.,~\cite{ren2017chartaccent, hullman2013contextifier, rahman2024qualitative}).
    \item Empirical studies on annotation usage that offer data-driven insights into how annotations affect user comprehension and engagement with visualizations in different contexts (e.g.,~\cite{stokes2022striking, borkin2015beyond}).
    \item Studies on tools and techniques for authoring visualizations that facilitate creating and integrating annotations in various usage contexts (e.g.,~\cite{ren2017chartaccent, chen2010click2annotate,isenberg2011collaborative, hullman2011visualization, groth2006provenance}).
\end{itemize}

\noindent
The exclusion criteria of our survey include:

\begin{itemize}
    \item Research focusing on text labeling and placement in Cartography (e.g.,~\cite{jones1989cartographic, christensen1995empirical}) because it deals with geographical data and spatial relationships, using methods distinct from those in non-spatial data visualizations.
    \item Studies on annotation usage in Scientific Visualization (SciVis) (e.g.,~\cite{loughlin1994annotation, gerum2017clickpoints}) because they involve volumetric and spatial data, which are fundamentally different from the abstract, non-spatial visualizations typical in information visualization.
    \item Studies on annotations in Immersive Analytics and AR/VR systems (e.g.,~\cite{danyluk2020touch, wither2009annotation}) because they focus on immersive environments with unique interaction paradigms, differing from traditional non-spatial visualizations. Recent surveys of these systems further reduce the need for inclusion (e.g.,~\cite{ARVRSurvey, IASurvey}).
    \item Research on data and image annotation in unrelated domains, such as Genomology (e.g.,~\cite{frishman2009modern,queiros2021unification}), Machine Learning (e.g.,~\cite{deng2009imagenet,lin2014microsoft}), and Library Science (e.g.,~\cite{svenonius2000intellectual, taylor2008organization}), etc., because they do not directly pertain to information visualization.
\end{itemize}

The distribution of selected papers by year is shown in \autoref{fig:papers_by_year}, and their breakdown by source is in \autoref{tab:source}.

\begin{table}[htb]
\centering
\caption{The Number of Surveyed Papers by Source}
\scriptsize %
\resizebox{\linewidth}{!}{%
\begin{tabular}{p{6.0cm}c}
\toprule
Sources & Count \\
\midrule
IEEE Trans. on Vis. and Comp. Graphics (TVCG) & 78 \\
IEEE Visualization and Visual Analytics (VIS) & \\
IEEE Information Visualization (InfoVis) & \\
IEEE Visual Analytics in Sci. and Tech. (VAST) & \\
\midrule
Computer Graphics Forum (CGF) & 14 \\
EG/IEEE VGTC Conf. on Visualization (EuroVis) & \\
\hspace{0.5cm} including Short Papers & \\
\midrule
ACM Conf. on Human Factors in Infor. Systems (CHI) & 50 \\
\hspace{0.5cm} including Extended Abstracts & \\
\midrule
Information Visualization Journal & 9 \\
\midrule
Visual Informatics & 8 \\
\midrule
Others—IEEE PacificVis; IEEE CG\&A; IEEE Computer; ACM IMWUT; ACM UIST; ACM Proc. on HCI; ACM ISS; ACM AVI; Comm. of the ACM; SimAUD; 
etc. & 32 \\
\bottomrule
\end{tabular}
}
\label{tab:source}
\end{table}

\section{Annotation Primer}
\label{sec.annotation-primer}

\subsection{What is an annotation?}
\label{sec.design-space.what}

Annotations have a long history, evolving from notes in manuscripts to printed books in the early modern period, providing commentary and cross-references~\cite{clanchy2012memory, blair2010too}. By the 17th and 18th centuries, scientists used annotations in lab notebooks to document experiments and validate results~\cite{shapin1995social}. The digital revolution expanded annotation use across disciplines: in software engineering, they serve as metadata to improve code readability~\cite{bloch2017effective, horvath2022using}; in bioinformatics, they describe functional aspects of genetic sequences~\cite{frishman2009modern, queiros2021unification}; in machine learning, they provide labeled data for model training~\cite{deng2009imagenet, lin2014microsoft}; in natural language processing, they mark text for linguistic analysis~\cite{marcus1993building, palmer2005proposition}; and in library science, they enhance bibliographic records for information retrieval~\cite{svenonius2000intellectual, taylor2008organization}.

Traditionally, annotations were primarily textual~\cite{cousins2000systems}, but their use has expanded to include graphical elements, especially in visualizations. In the visualization community, annotations are conceptualized differently: Munzner describes them as graphical or textual elements manually added to visualizations, becoming new attributes when linked to data items~\cite{munzner2014visualization}. Kong et al. define annotations as textual inputs like summary statistics and tooltips and graphical shapes such as rectangles and circles, serving as external cues that provide contextual information~\cite{kong2017internal}. In another study, Kong et al. highlighted annotations as graphical overlays that aid understanding without adding visual clutter, aligning with broader concepts like reference structures and highlights~\cite{kong2012graphical}. Chen et al.'s recent definition similarly frame annotations as graphic or textual additions that supplement or contextualize data, clarifying its significance in visualizations~\cite{chen2023does}.

Synthesizing these perspectives, Rahman et al.\cite{rahman2024qualitative} define annotation as: \textit{“Annotations are supplementary elements, such as graphical shapes, textual content, or adjusted colors, added to a visualization to enhance the basic data presentation by drawing attention to specific sections or aspects of the data.”} This definition includes supplementary elements that go beyond the basic components of a visualization, such as text (i.e., phrases, values, descriptions, etc.), graphical elements (i.e., rectangles, circles, lines, arrows, etc.), and color adjustments (highlights, color identifiers, etc.). A key aspect is distinguishing between basic elements and annotations in visualizations. Following Ren et al.'s approach\cite{ren2017chartaccent}, Rahman et al. did not consider direct data representations (e.g., bars in bar charts, scatterplot points, or map labels) as annotations. Likewise, essential chart components such as legends, tick marks, axes, etc., are excluded, as those elements are necessary for conveying information independently of annotations~\cite{rahman2024qualitative}.

\subsection{Design Spaces of Annotation}
\label{sec.design-spaces}
There are several design spaces of annotations in the visualization community, each tailored to specific aspects and applications. Hullman et al.\ examine 136 professionally created visualizations to categorize annotations into two types: additive annotations, which provide external context, and observational annotations, which highlight or emphasize existing data points and trends. They highlight the importance of anchoring annotations at various visual levels---single data points, groups or regions, and entire visualization views---to ensure relevance and engagement~\cite{hullman2013contextifier}. Ren et al.\ analyze 106 annotated charts, including bar charts, line charts, and scatterplots from professional sources, and identify distinct annotation forms---text, shapes, highlights, and images---which they apply in an interactive annotation tool named ChartAccent~\cite{ren2017chartaccent}. They categorize annotation targets into data items, structural chart elements, coordinate spaces, and prior annotations, facilitating context and emphasis within the chart.

\begin{figure*}[!t]
    \centering
    
    \includegraphics[width=0.85\linewidth]{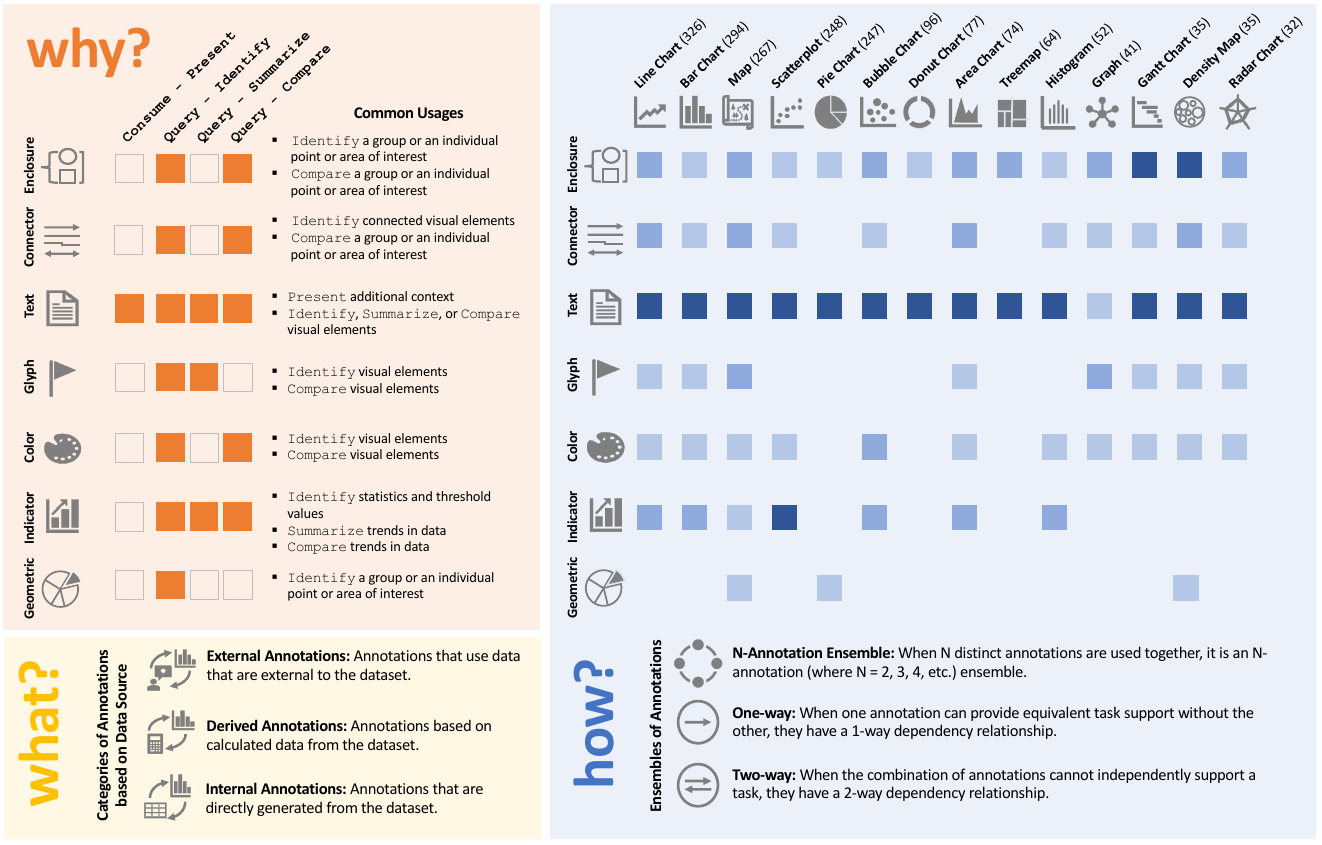}
    
    \caption{The design space by Rahman et al.~\cite{rahman2024qualitative} includes three sections: \taskWhy identifies visualization tasks and relevant annotation types, \taskHow details annotation usage with a frequency color-coding system: \fcolorbox{black}{dsLow}{\tiny 6-25\%}, \fcolorbox{black}{dsMed}{\tiny 26-50\%}, \fcolorbox{black}{dsHigh}{\tiny \textcolor{white}{51+\%}}, and types of annotation ensembles, and \taskWhat categorizes annotation data sources.}
    \label{fig:design-space}

\end{figure*}

While the previous two design spaces focused on the narrative aspect of professional visualizations, Rahman et al.\ proposed a comprehensive design space of annotations by examining about 1,800 static annotated charts of 14 common chart types from diverse data sources~\cite{rahman2024qualitative}. They developed a taxonomy that categorizes annotations based on their usage patterns and data sources and discusses the common analytic tasks they support based on the ``Why? How? What?" framework of Brehmer et al.~\cite{Brehmer2013} (see~\autoref{fig:design-space}). They also introduced annotation ensembles, which combine multiple types of annotations to convey complex information.

\subsection{Annotation Types}

Researchers have categorized annotations based on the two fundamental forms they take in visualizations: textual and graphical annotations (see \autoref{table:annotation-primer})~\cite{munzner2014visualization}.

\subsubsection{Textual Annotations}

\begin{figure*}[!t]
    \centering
    \includegraphics[width=0.8\linewidth]{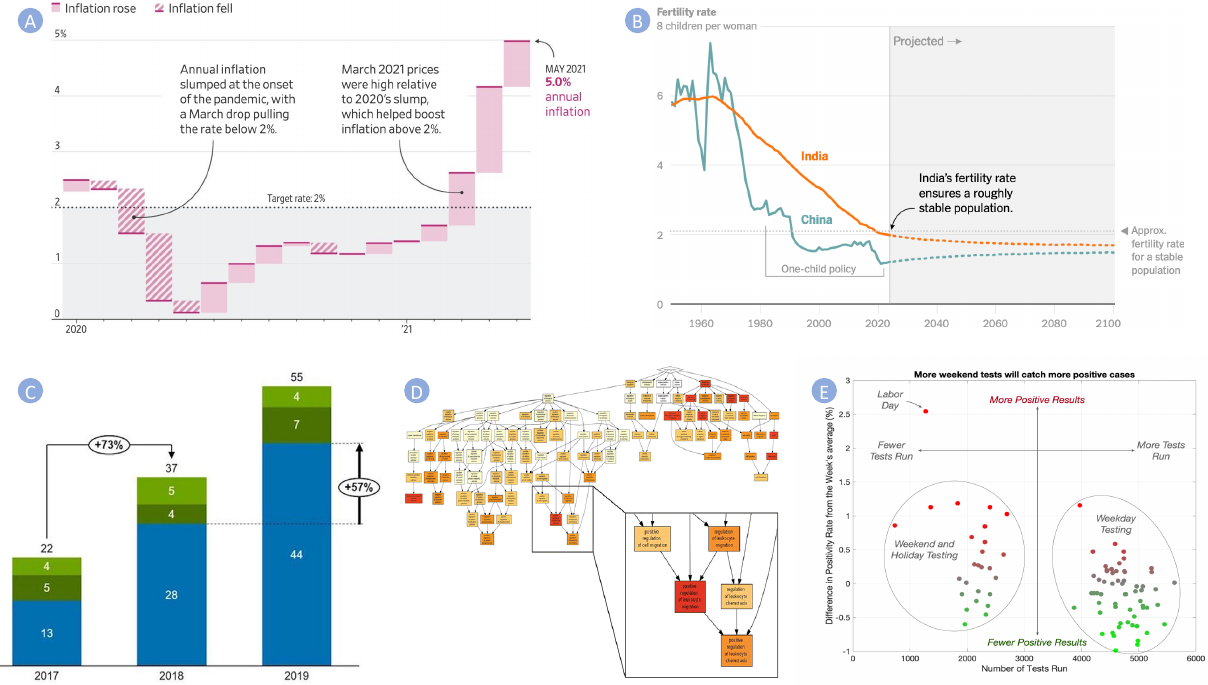}
    \caption{(A) and (B) are examples of professionally annotated charts: (A) A bar chart from \textit{The Wall Street Journal}~\cite{DapenaSantilli2021} highlighting inflation during COVID-19 with text, arrows, and highlighted region annotations; (B) A line chart from \textit{The New York Times}~\cite{TravelliCai2023} comparing fertility rates of India and China, use text annotations and highlights for emphasis. (C), (D), and (E) are collected from internet~\cite{rahman2024qualitative}: (C) A bar chart uses connectors, indicators, enclosures, and text annotations to facilitate comparison; (D) A node-link diagram uses enclosures and geometric annotations to identify areas of interest; (E) A scatterplot utilizes text, connectors, and enclosures to highlight data points.}
    \label{fig:examples}
\end{figure*}

Textual annotations involve using words, phrases or sentences to convey information about the data presented in visualizations~\cite{rahman2024qualitative}. Hullman et al.\ further classified textual annotations based on their functional roles in visualizations: additive and observational annotations~\cite{hullman2013contextifier}. Additive annotations enrich data visualizations by integrating external information, thus enhancing viewer understanding through background information, alternative viewpoints, or related events (see~\autoref{fig:examples}A-B). In contrast, observational annotations provide insights directly related to the displayed data, focusing on specific data aspects such as anomalies or notable variations (see~\autoref{fig:examples}C/E). For example, Bromley et al.'s work on semantic labeling of visual features in line charts enhances observational annotations by providing a structured approach to describing data characteristics such as trends and anomalies~\cite{bromley2023difference}. Additionally, their use of large language models (LLMs) for generating textual summaries supports additive annotations by incorporating external context and background information.

\begin{table}[!b]
\centering
\caption{Outline of annotation types (top), targets (middle), and generation methods (bottom).}
\label{table:annotation-primer}
\renewcommand{\arraystretch}{1.0} %
\setlength{\tabcolsep}{6pt} %
\begin{adjustbox}{max width=\linewidth}%
\arrayrulecolor[HTML]{000000}%
\begin{scriptsize}%
\begin{tabular}{|p{0.9cm}|p{1.3cm}|p{4.4cm}|p{1.0cm}|} %
\hline
\rowcolor[HTML]{EFEFEF} 
\textbf{Type} & \textbf{Subtype} & \textbf{Description} & \textbf{Examples} \\
\hline

\multirow{6}{*}{Textual} & 
\multirow{3}{*}{Additive} & 
Enrichment with external info, such as background details or related events (e.g., links to news articles) & 
\multirow{6}{1.2cm}{\cite{hullman2013contextifier}, \cite{wongsuphasawat2017voyager, bavoil2005vistrails, zhao2021chartstory}} \\
\cline{2-3}

 & 
\multirow{3}{*}{Observational} & 
Provide insights on displayed data, focusing on anomalies or notable variations (e.g., highlighting forecasting errors) & 
\\ 
\specialrule{.1em}{.05em}{.05em} 

\multirow{8}{*}{Graphical} & 
\multirow{2}{*}{Shapes} & 
Enclosures (rectangles, ellipses), connectors (lines, arrows), and marks (shape identifiers) & 
\multirow{8}{*}{\cite{kong2019understanding, zhao2020chartseer, wongsuphasawat2017visualizing, hulstein2022geo}} \\
\cline{2-3}

 & 
\multirow{3}{*}{Highlights} & 
Colors and shapes to emphasize data points, with adjustments to size, stroke, and external cues like contrast changes & 
\\ 
\cline{2-3}

 & 
Trends & 
Indicate data trends in visualizations & 
\\ 
\cline{2-3}

 & 
Images & 
Include icons and pictorial representations & 
\\ 
\cline{2-3}

 & 
Geometric & 
Enlarging or zooming into parts of a chart & 
\\ 
\hline

\end{tabular} 
\end{scriptsize}
\end{adjustbox}

\vspace{3pt}
\begin{adjustbox}{max width=\linewidth}%
\begin{scriptsize}%
\begin{tabular}{|p{2.3cm}|p{4.8cm}|p{1.0cm}|} %

\hline
\rowcolor[HTML]{EFEFEF} 
\textbf{Targets} & 
\textbf{Description} & 
\textbf{Examples} \\
\hline

Data item, set, \& series targets & 
Annotations on specific data points, series, or sets, emphasizing critical data or relationships & 
\multirow{8}{*}{\cite{ren2017chartaccent, hullman2013contextifier}} \\
\cline{1-2}

Coordinate space targets & 
Annotations within the chart's coordinate system, highlighting values and ranges for context & 
\\ 
\cline{1-2}

\multirow{2}{*}{Chart element targets} & 
Annotations on structural components like titles, axes, and legends, providing context or details & 
\\ 
\cline{1-2}

\multirow{2}{*}{Prior annotations} & 
New annotations added to existing ones, enhancing depth and interconnections & 
\\ 
\hline
\end{tabular} 
\end{scriptsize}
\end{adjustbox}

\vspace{3pt}
\begin{adjustbox}{max width=\linewidth}%
\begin{scriptsize}%
\begin{tabular}{|p{1.6cm}|p{4.9cm}|p{1.6cm}|} %

\hline
\rowcolor[HTML]{EFEFEF} 
\textbf{Generation} & 
\textbf{Description} & 
\textbf{Examples} \\
\hline

\multirow{2}{*}{Manual} & 
Annotations created manually using free-hand drawing, digital pens, or mice & 
\multirow{2}{*}{\cite{lee2015sketchinsight, bach2016telling, ellis2004collaborative, choi2015visdock}}
\\ 
\hline

Semi-automatic & 
Combines user input with predefined options &
\scriptsize{\cite{ren2017chartaccent, xu2018chart, wood2018rethinking, stolper2016emerging}}
\\ 
\hline

\multirow{2}{*}{Automatic} & 
Algorithm-driven annotations, highlighting key features without user input & 
\scriptsize{\cite{hullman2013contextifier, kandogan2012just, ibanez2023almanac, zhao2016annotation}}
\\ 
\hline

\end{tabular}
\end{scriptsize}
\end{adjustbox}
\end{table}

\subsubsection{Graphical Annotations}

The other fundamental form of annotation is graphical, which broadly encompasses modifications to the visual elements of a visualization aimed at emphasizing or drawing attention to specific sections. Graphical annotations can take many different forms in visualizations according to the existing literature. For example, shapes include enclosures (e.g., rectangles, ellipses) (see~\autoref{fig:examples}C-E), connectors (e.g., lines, arrows) (see~\autoref{fig:examples}A-B), and marks (e.g., different shape identifiers)~\cite{ren2017chartaccent, rahman2024qualitative}. Highlights involve utilizing various colors and shapes to emphasize or de-emphasize data points, with adjustments in size and stroke, and also include color for highlighting and external cues such as contrast-based changes to emphasize focus areas or dim surrounding contexts~\cite{ren2017chartaccent, rahman2024qualitative} (see~\autoref{fig:examples}A-B). Images encompass icons and other pictorial representations~\cite{ren2017chartaccent}. Trends indicate data trends in visualizations. Finally, geometric annotations involve enlarging or zooming into parts of a chart~\cite{rahman2024qualitative} (see~\autoref{fig:examples}D).

Beyond static textual or graphical additions, Heer et al.\ introduced data-aware annotations as dynamic elements, both textual and graphical, that react to interactions with data, such as selection or brushing~\cite{heer2012interactive, heer2008generalized}. This is evident in tools such as Click2Annotate~\cite{chen2010click2annotate} and Touch2Annotate~\cite{chen2010touch2annotate}, where annotations are generated in response to user interactions with data points, ensuring that they are contextually relevant and triggered by specific user actions. Data-aware annotations are valuable for anchoring annotations to specific data points and facilitating the reuse of annotations across different visual representations~\cite{heer2012interactive}.

When a single annotation is insufficient to denote a phenomenon, multiple annotation types are combined, which Rahman et al.\ defined as ensembles of annotations in their work~\cite{rahman2024qualitative, rahman2022qualitative, rahman2023exploring, rahman2024exploring}. For example, in~\autoref{fig:examples}A-B), lines and arrows connect text descriptions to specific points of interest on the charts. In this case, connectors and textual annotations work together to denote a single phenomenon, forming an ensemble of annotations. Ren et al.\ also discussed using multiple annotations for a single target in ChartAccent~\cite{ren2017chartaccent}.

\subsection{Annotation Targets}
 Annotation targets are elements in a visualization that can be highlighted to provide additional information, context, or emphasis. These targets serve as key components for designing annotation authoring tools, ensuring that annotations are effectively integrated into visualizations. Ren et al.\ identified four primary annotation targets for visualizations: \textit{Data Item, Set, \& Series Targets}, \textit{Coordinate Space Targets}, \textit{Chart Element Targets}, and \textit{Prior Annotations} (see \autoref{table:annotation-primer})~\cite{ren2017chartaccent}.

\begin{figure}[!t]
    \centering
    {\includegraphics[trim=0 0 350pt 0, clip, width=0.85\linewidth]{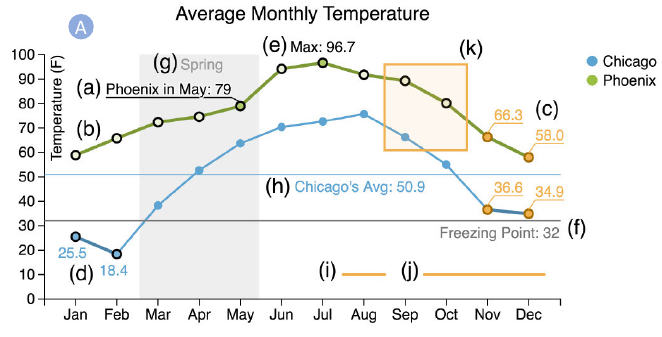}}
    {\includegraphics[trim=330pt 0 0 0, clip, width=0.95\linewidth]{chartaccent.pdf}}
    \caption{(A) Chart of average monthly temperatures with various annotation targets from Ren et al.~\cite{ren2017chartaccent}. (B) ChartAccent interface: (1) manual annotation area; (2) highlight annotation; (3) control panel; (4) annotation~list; (5) editor; (6) target selection; (7) visual adjustment~tools.}
    \label{fig:chartaccent}
\end{figure}

\autoref{fig:chartaccent} shows different annotation targets from Ren et al.'s study. \textit{Data item, set, \& series targets} encompass annotations on specific data points, entire data series, or sets of items. They emphasize critical data, illustrate relationships, or guide viewers through the chart's structure. Examples include highlighting a particular temperature or comparing data across different categories (see~\autoref{fig:chartaccent}A: a-e). \textit{Coordinate space targets} pertain to annotations within the chart's coordinate system. They might highlight specific values and ranges or span across dimensions, aiding in contextualizing and orienting the viewer within the chart's framework (see~\autoref{fig:chartaccent}A: f-k). \textit{Chart element targets} refer to annotations on the structural components of the chart itself, such as titles, axes, and legends. These annotations often provide broader context or explanatory details about the data and the chart's methodology. \textit{Prior annotations} represent a meta-level, where new annotations are added to existing ones. This layering approach enhances the depth of information and interconnections within the chart, allowing for more nuanced interpretation and reference (see~\autoref{fig:chartaccent}A: a-k). Hullman et al.\ in Contextifier~\cite{hullman2013contextifier} also discussed annotation targets, focusing on single data points, groups of data, and entire visualizations, closely resembling the targets identified by Ren et al.

\subsection{Annotation Generation Methods}

We categorized the generation of annotations into three fundamental approaches: manual, semi-automatic, and automatic, each with different levels of user interaction and automation (see \autoref{table:annotation-primer}).

\subsubsection{Manual}
\begin{figure}[!b]
  \centering
  \includegraphics[width=0.9\linewidth]{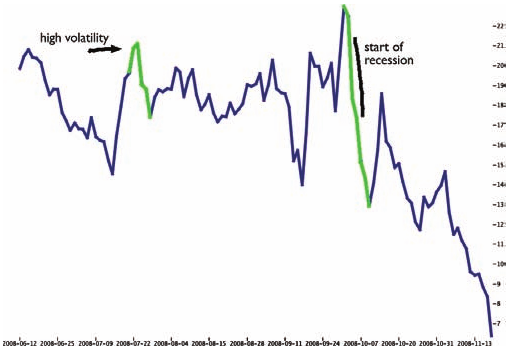} 
  \caption{Stock chart with user-drawn annotations: an arrow highlighting a peak and a line showing a decline~\cite{kong2009perceptual}. The algorithm interprets these ink annotations, identifying key chart features (marked in green) like peaks and slopes, enhancing interpretation in visual discussions.}
  \label{fig:InkAnnotation}
\end{figure}

Manual methods involve direct user input for creating annotations, with two primary tools being used: digital pens and mice. The digital pen method, exemplified by tools such as SketchInsight~\cite{lee2015sketchinsight}, PanoramicData~\cite{zgraggen2014panoramicdata}, TimeSplines~\cite{offenwanger2023timesplines}, Newsr~\cite{wood2018rethinking} and ActiveInk~\cite{romat2019activeink}, allows users to draw free-form annotations directly on the visualizations, offering a high degree of personalization (see~\autoref{fig:InkAnnotation}). On the other hand, the mouse-based approach, as utilized in tools developed by Heer et al.~\cite{heer2007voyagers} and Denisovich et al.~\cite{denisovich2005software}, enables users to create annotations using a more traditional computer interface. Both methods provide a highly user-driven experience in the annotation process.

\subsubsection{Semi-automatic} 
The semi-automatic approach combines user-driven inputs with predefined annotation options. This method typically features an annotation palette from which users can select predefined annotation types. ChartAccent~\cite{ren2017chartaccent}, for example, allows users to choose from a range of annotation types (i.e., text boxes, lines or arrows, enclosures) within an annotation palette (see~\autoref{fig:chartaccent}B). Similarly, commercial data visualization tools such as Tableau~\cite{tableau}, Microsoft Power BI~\cite{microsoft_power_bi}, and Google Data Studio~\cite{google_data_studio} provide users with options to choose their preferred annotations from a selection of predefined types (more discussions on these in~\autoref{general_purpose_tools}). Additionally, some tools in this category enable interactive annotation by allowing users to click (e.g., Click2Annotate~\cite{chen2010click2annotate}) or touch (e.g., Touch2Annotate~\cite{chen2010touch2annotate})  items in the visualizations, applying graphical and textual annotations. This approach balances manual control with ease of use. Most of these tools also allow users to customize their annotations by changing their size, style, position, etc.

\subsubsection{Automatic}
\begin{figure*}
    \centering
    \includegraphics[width=0.9\linewidth]{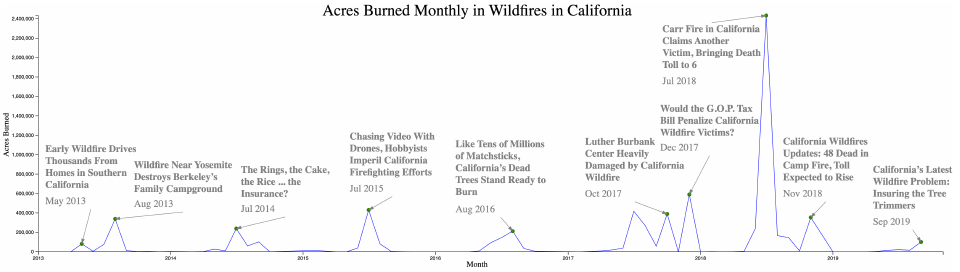}
    \caption{Automatically generated annotations on a time-series chart of California wildfires (2013–2020) using the Almanac API~\cite{ibanez2023almanac}, highlighting key wildfire events.}
    \label{fig:AutomaticAnnotation}
\end{figure*}

Automatic methods involve no user input in the annotation process, relying instead on algorithms to identify and highlight key features of the visualizations. Contextifier~\cite{hullman2013contextifier}, for instance, automatically generates annotations for stock visualizations by integrating linguistic relevance from news articles with visual salience from stock data. Similarly, the system developed by Kandogan et al.\cite{kandogan2012just} identifies data elements such as clusters, outliers, and trends and automatically applies textual annotations, facilitating an enhanced understanding of complex datasets without direct user involvement. Some tools let the users interactively adjust the automatically generated annotations\cite{bryan2016temporal, walker2015storyboarding}. \cite{ibanez2023almanac, bryan2016temporal}, \cite{kong2009perceptual, latif2021kori, bromley2023difference, wongsuphasawat2017visualizing, lai2020automatic} are other studies that discuss automating the annotation process (see~\autoref{fig:AutomaticAnnotation}).

An interactive website classifying the surveyed papers based on different categories can be found at \websiteURL.

\section{Empirical Studies}
\label{sec.empirical-studies}

We explored empirical studies on annotations and discovered their many roles, such as improving data comprehension, enhancing memorability and recall, and supporting collaborative visual analysis and storytelling. Additionally, we surveyed the effect of annotations on user engagement and interaction, and how they can be used to mislead viewers (see \autoref{table:empirical-studies}).

\subsection{Comprehension, Memorability, and Recall}

Research in visualization literature demonstrates that annotations enhance data visualizations' comprehension, memorability, and recall. By capturing attention and aiding memory encoding, annotations make visualizations more memorable and informative~\cite{borkin2013makes, borkin2015beyond}. Clear headlines and color highlights direct focus and improve memory recall by reducing visual clutter~\cite{ajani2021declutter}.

\begin{table}[!t]
\centering
\caption{Summary of annotation empirical studies.}
\renewcommand{\arraystretch}{0.9} %
\setlength{\tabcolsep}{6pt} %
\begin{adjustbox}{max width=\linewidth} %
\arrayrulecolor[HTML]{000000} %
\begin{tabular}{|>{\columncolor[HTML]{EFEFEF}}p{1.75cm}|p{4cm}|p{2cm}|}
\hline
\rowcolor[HTML]{EFEFEF} 
\scriptsize \textbf{Study Type} & 
\scriptsize \textbf{Descriptions} & 
\scriptsize \textbf{Examples} \\ 
\hline

\scriptsize Comprehension, Memorability, and Recall & 
\scriptsize Annotations enhance comprehension, memorability, and recall by reducing cognitive load and providing context. & 
\scriptsize{\cite{borkin2015beyond}, \cite{ajani2021declutter}, \cite{arunkumar2023image}, \cite{ottley2019curious}, \cite{stokes2022striking}, \cite{stokes2023role}, \cite{fan2024understanding}} \\ 
\hline

\scriptsize Collaboration & 
\scriptsize Annotations improve synthesis, communication, and decision-making in collaborative environments. & 
\scriptsize{\cite{robinson2008collaborative}, \cite{chen2011supporting}, \cite{kadivar2009capturing}, \cite{bresciani2009benefits}, \cite{mahyar2012note}} \\ 
\hline

\scriptsize Narrative Storytelling & 
\scriptsize Annotations add depth and clarity, enhance narrative coherence, and improve engagement in visual storytelling. & 
\scriptsize{\cite{kosara2013storytelling}, \cite{boy2015storytelling}, \cite{segel2010narrative}, \cite{hullman2013deeper}, \cite{stolper2016emerging}, \cite{wang2019comparing}, \cite{bach2016telling}, \cite{shi2021communicating}} \\ 
\hline

\scriptsize User Engagement and Interaction & 
\scriptsize Annotations guide attention, facilitate diverse interpretations, and enrich user interaction and engagement. & 
\scriptsize{\cite{kong2017internal}, \cite{heer2008graphical}, \cite{kauer2021public}, \cite{sauve2022put}, \cite{taher2015exploring}, \cite{wood2012sketchy}, \cite{kim2019inking}, \cite{moore2021exploring}} \\ 
\hline

\scriptsize Misleading Viewers & 
\scriptsize Annotations can mislead viewers, highlighting the need for accurate, context-rich annotations. & 
\scriptsize{\cite{lisnic2023misleading}, \cite{ge2023calvi}, \cite{ritchie2019lie}, \cite{song2018s}} \\ 
\hline
\end{tabular}
\end{adjustbox}
\label{table:empirical-studies}
\end{table}

Textual annotations play an important role in improving data comprehension in visualizations by reducing cognitive load and enhancing interpretive accuracy. Research has shown that visualizations highlight key information while text aids in understanding it, making the integration of text and visuals crucial for effective information processing~\cite{ottley2019curious, zhi2019linking}. Combining text and visualizations reduces cognitive load and improves comprehension~\cite{bancilhon2023combining, 6327259, bryan2020analyzing}, whereas the absence of annotations increases the likelihood of misinterpretation or a less nuanced understanding of the data~\cite{koesten2023message}, confirming the value of this approach. Furthermore, text annotations improve visualization comprehension, with simpler labels promoting reliance on the visual elements and more detailed annotations offering deeper insights, highlighting the importance of aligning annotation detail with the level of visual data granularity~\cite{fan2024understanding}. User preferences tend to favor heavily annotated charts, which aid in comprehension and influence viewers' perceptions and predictions~\cite{stokes2022striking, stokes2023role, arunkumar2023image}. Textual annotations are also essential for guiding novices, those less familiar with visualizations, by providing context, clarifying insights, and aiding the interpretation of complex visuals. Although not directly annotations, similar studies have focused on the critical linkage between text and visualizations~\cite{kong2019trust, burns2022invisible, kim2021towards, kong2018frames}.

The effectiveness of annotations varies based on their type and context~\cite{guo2024we}, with different types of annotations and visual enhancements contributing uniquely to the effectiveness of data visualizations. Additive annotations that introduce new content have been found to improve comprehension, whereas observational annotations may not enhance understanding more than unannotated visualizations~\cite{chun2020giving}. The combination of visual cues with audio narrations can also impact comprehension and recall, though this effect depends on the specific techniques and chart types employed~\cite{kong2019understanding}. Symbols and pictograms have been shown to enhance both comprehension and recall in visual note-taking~\cite{zheng2021sketchnote}. Moreover, textual annotations that are automatically linked to visual elements using a knowledge graph, and then refined by users, can improve the accuracy and effectiveness of the visualizations~\cite{cai2024linking}. Interestingly, visual embellishments, often referred to as `chartjunk,' do not impair accuracy and can enhance long-term recall, indicating that additional visual elements can enrich viewer comprehension and retention similarly to annotations~\cite{bateman2010useful}.

\begin{figure}[!b]
    \centering
    {\includegraphics[trim=0 0 237pt 0, clip, width=0.9\linewidth]{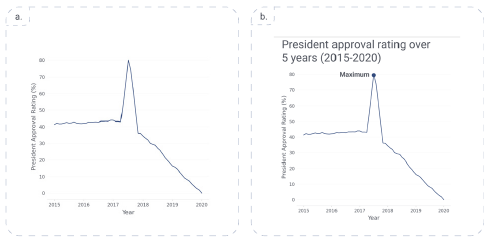}}
    {\includegraphics[trim=237pt 0 0 0, clip, width=0.9\linewidth]{StrikingABalance.pdf}}
    \caption{Stokes et al.~\cite{stokes2022striking} examined the impact of varying text quantity on visualization effectiveness in line charts.}
    \label{fig:StrikingABalance}
\end{figure}

\subsection{Impact in Narrative Storytelling}
Annotations enhance the storytelling aspect of data visualizations by providing depth and clarity and facilitating smooth transitions. They enable interactive exploration across various mediums and attract viewer attention during observation and recall tasks in narrative visualizations, guiding perception and shaping how users process and understand visual information~\cite{bryan2020analyzing}. Strategic use of annotations is crucial for maintaining narrative coherence and preventing viewer overload. Balanced integration of textual and graphical annotations helps avoid confusion in complex storytelling scenarios, guiding viewers and setting the stage for interactive exploration~\cite{kosara2013storytelling, boy2015storytelling}. Annotations guide attention and clarify information, enhancing comprehension and engagement across various contexts in narrative visualizations~\cite{zhi2019linking}. In online journalism, they prevent misunderstandings and maintain reader interest~\cite{segel2010narrative, haroz2015connected}, whereas in professional visualizations, elements such as labels, titles, and tooltips ensure smooth narrative transitions, guiding viewer attention and improving understanding~\cite{hullman2013deeper, stolper2016emerging}.

Annotations also play a significant role in various narrative storytelling mediums. For example, annotations reduce redundancy in data comics and improve text-picture integration~\cite{wang2019comparing}. In graph comics, they significantly improve comprehension of dynamic network changes~\cite{bach2016telling}. Data videos and visualization thumbnails for journalism benefit from annotations that enhance data comprehension, narrative cohesion, and attractiveness~\cite{shi2021communicating, 8933773}. Video presentations incorporating personal stories become more engaging and relatable with well-placed annotations~\cite{choe2015characterizing}. Additionally, text highlighting methods can enhance or hinder text analysis, emphasizing the importance of well-designed annotation techniques~\cite{strobelt2015guidelines}.

\subsection{Impact in Collaboration}

Annotations support collaborative visual data analysis by externalizing insights, enhancing communication, integrating complex datasets, and emphasizing key data points, which improves group performance~\cite{mahyar2012note, robinson2008collaborative, kang2014characterizing}. They help experts bridge communication gaps and make informed decisions by clearly presenting critical information. In asynchronous visual analytics, annotations allow team members to add and view comments and insights over time, improving collective understanding and decision-making~\cite{chen2011supporting}. Hand-drawn annotations assist in collaborative discovery by visually marking patterns and objects that might otherwise go unnoticed, encouraging deeper engagement among team members~\cite{kadivar2009capturing, wood2018rethinking}. In organizational and co-located settings, annotations enhance group performance and knowledge sharing by making key information accessible and understandable, facilitating more effective collaboration~\cite{bresciani2009benefits, mahyar2012note}.

\subsection{Increasing User Engagement and Interaction}
Annotations play a key role in enhancing user engagement and interaction with data visualizations by guiding attention, clarifying information, and supporting exploration. Studies show that annotations, such as threshold lines, shaded regions, and discursive patinas, help users highlight key data points, connect discussions to visual elements, and provide context, leading to improved understanding and interaction with visualizations~\cite{choi2015visdock, kauer2024discursive, guo2024we}. Techniques that combine annotations with internal cues, like brightness or transparency adjustments, further enhance comprehension and audience engagement~\cite{kong2017internal}. Additionally, annotations such as bookmarking and keyword tagging facilitate meaningful interaction by adding context to visual data~\cite{heer2008graphical}.

In shared platforms, user-generated annotations enhance engagement by supporting diverse data interpretations through shared observations, critiques, and insights~\cite{kauer2021public}. Different rendering styles and digital annotation tools further enrich user engagement and interaction. For example, sketchy rendering styles can influence user annotations by encouraging exploratory interactions with data~\cite{wood2012sketchy}. In personal informatics, annotations add context and personal insights, turning raw data into meaningful narratives and enhancing user engagement~\cite{moore2021exploring}. Dynamic graphical annotations boost user engagement by enabling users to interact directly with data, highlight key points, and visually track their analysis, making data exploration more interactive and focused~\cite{kim2019inking, taher2015exploring}.

\subsection{Misleading Viewers}

Research reveals that annotations can sometimes mislead viewers, making it crucial to design them carefully to ensure accurate data interpretation. Deceptive annotations in COVID-19 visualizations on social media—such as omitting crucial information or attributing data changes to unrelated causes—lead to false interpretations and highlight the need for accurate, context-rich annotations~\cite{lisnic2023misleading}. Misleading annotations are major contributors to visualization misinformation, even though they might be easy to spot~\cite{ge2023calvi}. To tackle this issue, techniques such as `Perceptual Glimpses' have been proposed, which switch between dataset views to reveal critical trends and reduce misleading impacts. This approach relies on clear annotations to help viewers correctly interpret the data~\cite{ritchie2019lie}. Additionally, highlighting and annotating imputed values with statistical information can greatly improve perceived data quality and accuracy, whereas downplaying or omitting information degrades these perceptions. Studies show that participants prefer visualizations with clear annotations indicating missing data, as this increases the credibility and confidence in the visualized information~\cite{song2018s}. Also, annotations, such as text explanations and labels, are essential for clarifying complex data and making it more accessible and trustworthy to the public~\cite{yang2023swaying}.

\section{Applications and Utilities}
\label{sec.application}

We focus on tools, techniques, and frameworks for annotations, covering general-purpose tools and specialized approaches using annotations in narrative visualizations, collaborative analytics, exploratory data analysis, provenance tracking, uncertainty visualizations, and user interaction. The goal is to demonstrate how annotations are applied across platforms to support data comprehension and analysis in different contexts (see \autoref{table:application-annotation}).

\subsection{General Purpose Tools and Visualization Libraries}
\label{general_purpose_tools}

Over the past two decades, several commercial data visualization tools have become widely used. These tools, such as Tableau~\cite{tableau}, Microsoft Power BI~\cite{microsoft_power_bi}, and Google Data Studio~\cite{google_data_studio}, commonly emphasize improving user interaction and comprehension through various annotation features.

Tableau and Power BI both support extensive dashboard annotations, including textual and graphical elements as well as interactive tooltips, as demonstrated in their respective public forums~\cite{MicrosoftFabricCommunity2024, TableauPublicDiscover, TableauIronViz}. Tableau enables embedding customizable text directly within visualizations and using elements such as arrows, lines, and shapes to visually link and emphasize different parts of the data. Power BI focuses more on textual annotations through customizable text boxes and uses various visualization types or external custom visuals for graphical annotations. Its interactive tooltips aim to present data comprehensively without cluttering the visual space. Google Data Studio, less popular than Tableau and Power BI, adopts a straightforward approach to annotations, enabling annotations through text boxes and shapes and dynamic content creation via custom-calculated fields. Its interactive elements provide detailed hover text, facilitating a clearer presentation.

The annotation capabilities of D3.js and Vega, two leading web-based visualization libraries, demonstrate differing approaches. D3.js offers detailed control over SVG elements, providing a highly customizable platform for annotations. Users can manipulate all aspects of an annotation, including positioning, shape, style, and interactivity, such as hover or click responses. Tools such as `d3-annotation'~\cite{lu_d3-annotation}, `swoopydrag'\cite{swoopyDrag}, and `labella.js'\cite{labella.js} enhance this flexibility, making D3.js ideal for developers who require tailored annotation solutions. On the other hand, Vega employs a high-level, JSON-based format that simplifies the creation and styling of annotations using built-in marks such as `text' and `rule'. This method reduces coding complexity and is more accessible to those less versed in SVG or JavaScript. Although Vega supports interactive annotations, its structured approach offers less granularity than D3.js. Therefore, D3.js caters to developers seeking extensive customization, whereas Vega suits those who prefer a more straightforward method for standard annotation tasks in visualizations.

Web-based visualization libraries implement several strategies. VisDock~\cite{choi2015visdock} treats annotations as independent overlays, promoting flexibility and collaboration in interactive web environments. In contrast, Protovis~\cite{2009-protovis} employs a declarative JavaScript approach, facilitating dynamic interactions and shared properties among visual elements. Matplotlib~\cite{tosi2009matplotlib} provides Python users with comprehensive annotation tools such as text, shapes, arrows, and interactive features. Conversely, ggplot2~\cite{wickham2011ggplot2} serves the R community by incorporating annotations through a layering system, which preserves the clarity of underlying data visualization. This method supports the addition of text labels, geometric shapes, and reference lines without disrupting the data display.

In summary, the variation in annotation strategies across visualization platforms reflects their tailored approaches to meet user needs. Commercial tools such as Tableau and Power BI provide basic and intuitive annotation features, such as text boxes and interactive tooltips, aimed at general users seeking simplicity and clarity. In contrast, specialized libraries such as D3.js cater to developers needing detailed customization options, allowing precise control over annotations' appearance and behavior. This differentiation underscores the importance of annotations in improving user interaction and understanding across different contexts, balancing functionality and usability to support various technical skills and analytical needs.

\subsection{Storytelling and Narrative Visualizations}

Narrative storytelling in data visualization uses engaging and explanatory narratives to present complex data effectively. Annotations enhance these narratives by providing context and clarity, improving interpretability and accessibility \cite{hullman2011visualization, martinez2020data, wohlfart2007story}.

Research on automated annotation techniques in narrative visualizations has focused on improving accessibility, interaction, and information delivery in data visualizations. Tools such as Contextifier and Almanac automate textual annotations for financial and time-series data, helping users understand trends through relevant visual cues and data analysis \cite{hullman2013contextifier, ibanez2023almanac}. ChartStory uses language stitching to generate automatic annotations in a comic-style format, aiming to make data presentations more engaging \cite{zhao2021chartstory}. Calliope-Net identifies and annotates complex graph data to make it more readable for non-expert users \cite{chen2023calliope}. Lai et al. developed a method to align textual descriptions with visual data points, improving the coherence between narrative and visuals \cite{lai2020automatic}. A recent approach employs reinforcement learning to automate annotations on animated scatterplots, sequentially highlighting key insights to improve understanding \cite{shi2024understanding}. The TSI tool generates relevant annotations automatically and allows users to adjust them interactively~\cite{bryan2016temporal}. Tools such as ChartAccent and Kori also combine automated and manual inputs to modify annotations based on interactions, drawing attention to specific data features for tailored narratives \cite{ren2017chartaccent, latif2021kori}.

Interactive visualization tools have been developed to improve storytelling through the use of annotations. For example, SketchStory allows presenters to create data charts through freeform sketching, supporting dynamic data filtering and annotations to make presentations more interactive \cite{lee2013sketchstory}. Ellipsis integrates storytelling with dynamic annotations in a graphical interface, supporting both linear and non-linear narratives \cite{satyanarayan2014authoring}. Narvis and Charagraphs use interactive annotations to explain complex visualizations and merge interactive charts with traditional texts, improving the reading experience \cite{wang2018narvis, masson2023charagraph}. Interactive timeline tools, such as Timeline Storyteller, NewsViews, and TimeSplines, provide annotations and flexible data manipulation for timelines and news analysis \cite{brehmer2019timeline, gao2014newsviews, offenwanger2023timesplines}. The storyboarding tool epSpread allows users to interactively add and adjust textual and graphical elements on storyboard panels, supporting the creation and refinement of visual narratives \cite{walker2015storyboarding}. Additionally, some tools and frameworks facilitate automatic annotation generation while also allowing users to manually and interactively adjust annotations \cite{bryan2016temporal, ren2017chartaccent, latif2021kori, lu2017visual}.

Annotations play a key role in storytelling with data comics, turning complex visualizations into narratives by making static data more interactive and understandable, articulating trends, explaining behaviors, and adding context to the storytelling process \cite{bach2018design, bach2016telling, zhao2015data, wang2020cheat}. DataToon supports the creation of data comics, offering annotations such as text labels, data group highlighting, and custom graphical elements through pen and touch methods, allowing users to add text labels for interactivity and personalization~\cite{kim2019datatoon}. ChartStory helps arrange charts into comic-style layouts, generates textual annotations to explain insights, and allows users to refine the narrative interactively. Another study introduced design patterns for data comics, offering a structured approach to using textual and graphical annotations for storytelling and data communication~\cite{bach2018design}.

In summary, annotations improve visual storytelling by providing clarity and context, making complex data more accessible. Research has focused on enhancing both automated and manual annotation techniques that adapt to user interactions and highlight specific data features in narrative visualizations. Innovations such as automated language stitching and contextually relevant annotations combine textual descriptions with visual data points, improving the narrative flow and effectiveness of presentations. These annotation techniques also improve data comics by making visualizations clearer and more interactive, aiming to make complex data easier to understand and engage with for a wider audience.

\subsection{Collaborative Visualizations}
\label{sec.collab}

Annotations are essential for enhancing interactivity and clarity in collaborative visual analysis and sensemaking. They serve a vital role in both synchronous and asynchronous communication within visualization systems, helping users to highlight key data points, share insights, and develop collective understanding~\cite{viegas2006communication, heer2007design, heer2007voyagers, kong2009perceptual, zhao2017supporting, wongsuphasawat2017voyager, willett2011commentspace}.

Research has developed tools that facilitate asynchronous collaborative data analysis through annotations, supporting both group and individual workflows. Many Eyes, for example, uses real-time and asynchronous annotations to make data visualization more accessible, incorporating textual annotations and shared visualization states to promote community engagement and information sharing \cite{viegas2006communication, viegas2007manyeyes}. Tools such as epSpread, Commentspace, Chart Constellations, and Voyager 2 enable users to add textual annotations directly on visualizations, panels, and bookmarked views, which aids in documentation, organized sharing, and collaborative exploration across different timescales \cite{walker2015storyboarding, willett2011commentspace, xu2018chart, wongsuphasawat2017voyager}. The Sandbox in nSpace supports manual graphical and textual annotations to document reasoning and organize evidence, facilitating both asynchronous and limited synchronous collaboration \cite{wright2006sandbox}. Parts-based segmentation algorithms interpret freeform ink annotations on line charts to highlight perceptual features, reducing ambiguity in asynchronous settings \cite{kong2009perceptual}. Systems such as the Knowledge-Transfer Graph (KTGraph) combine graph elements and interactive timelines with annotations, making it easier to share and document analytical processes asynchronously \cite{zhao2017supporting}. Another tool allows data-aware textual annotations linked to selection queries, enabling dynamic reuse across visual contexts and enhancing interactive data exploration in asynchronous environments \cite{heer2008generalized}.

Annotations support team-based collaborative data analysis and sensemaking in synchronous settings. Techniques such as embedding markers and linking comments within visual analytics environments are used to improve team interactions, as seen in platforms that integrate view sharing, graphical annotation, and social navigation \cite{heer2007design, heer2007voyagers}. Systems that support team collaboration via annotations include digital ink, voice, text, and deixis-centered frameworks that transform verbal and gestural interactions into synchronized textual and graphical annotations, enhancing data discussions in real-time collaborative environments~\cite{ellis2004collaborative, eccles2008stories, wongsuphasawat2017voyager, biehl2007fastdash, han2024deixis}. Tools such as VisConnect streamline the annotation process for distributed teams with dual-layer structures and integrated chat, whereas InsideInsights and SketchInsight enable collaborative decision-making and synchronous data analysis through manual and dynamic annotation features \cite{schwab2020visconnect, mathisen2019insideinsights, lee2015sketchinsight}.

In summary, annotations support interactivity and clarity in collaborative visual analysis and sensemaking, playing a key role in communication. In asynchronous settings, annotations help document insights, organize information, and facilitate collaborative exploration across different timescales. In synchronous environments, annotations enhance real-time team interactions by integrating various forms of input into graphical and textual annotations, improving communication and data discussions. Although annotations aid in highlighting key data points and sharing insights, they may also potentially introduce variability in interpretation and require careful design considerations to maintain effective communication across different contexts and platforms.

\subsection{Exploratory Data Analysis}
\label{sec.exploratory-analysis}

In exploratory data analysis, tools integrate textual and graphical elements to improve understanding of complex datasets. For instance, KronoMiner and Annotated TimeTree use interactive labels and color-coded markers to manage and explore time-series data and document collections, whereas VisInReport combines textual and graphical annotations to help users interpret complex information~\cite{zhao2011kronominer, xia2014annotatedtimetree, sevastjanova2021visinreport}. In discourse analysis, tools provide manual textual annotations through integrated panels, allowing researchers to document insights and refine parsing algorithms within the visualization environment~\cite{zhao2012facilitating}. For error detection, some tools use in situ annotations to identify and correct visual inconsistencies, functioning similarly to spell-check for charts~\cite{Arlene2022annotating, hopkins2020visualint}. Visual analytic tools employing dynamic color-coded updates clarify relationships between visual encoding and layout, aiding data interpretation~\cite{hoffswell2016visual}. In complex visualizations, such as large-scale graphs and matrix-based analyses, tools such as DataMeadow and MatrixExplorer use annotations to highlight key data points, uncover patterns, and support data organization and collaborative exploration~\cite{abello2006ask, elmqvist2008datameadow, henry2006matrixexplorer}.

Annotations help reduce cognitive load in data analysis by highlighting key features such as clusters and outliers with optimized fonts, making data interpretation more straightforward~\cite{kandogan2012just}. VisInReport integrates textual and visual annotations to connect diverse data types in discourse analysis, which aids in user understanding~\cite{sevastjanova2021visinreport}. VisJockey uses highlights, animations, and inset graphics to guide users through interactive visualizations, making complex data narratives easier to follow~\cite{kwon2014visjockey}. Another framework combines textual and graphical annotations, enabling users to explore data through contextual markups and interactive elements, helping to manage cognitive load and maintain engagement~\cite{latif2018exploring}.

Annotations support exploratory data analysis by allowing direct interaction with data, identifying patterns, and extracting insights from complex visualizations. In crime analysis, they enable interactive marking and commenting on maps, making spatial data accessible to non-technical users~\cite{de2020visual}. Jigsaw and ChartSeer use annotations to uncover hidden narratives and highlight trends to assist in the quick identification of patterns~\cite{stasko2008jigsaw, zhao2020chartseer}. Integrating annotations with computational features supports detailed exploration in complex scenarios, such as multi-agent simulations~\cite{breslav2015exploratory}, whereas the Irvine system enables manual annotations on spectrograms, helping engineers document faults and understand manufacturing issues in electrical engines~\cite{eirich2021irvine}.

In summary, annotations assist in exploratory data analysis by supporting user interaction with complex datasets, helping to identify critical features, document insights, and correct visual inconsistencies. Tools incorporating textual and graphical annotations aid in interpreting information, exploring data relationships, and managing cognitive load. By facilitating direct interaction with data, annotations can help users uncover patterns, refine analysis processes, and stay engaged. However, thoughtful integration of annotations is key to aiding the exploratory process and engaging users without overwhelming them.

\subsection{Provenance Data Visualizations}
\label{sec.provenance}

\begin{table*}[!ht]
\centering
\caption{Annotation Types and Generation Methods by Application Area}
\renewcommand{\arraystretch}{0.9} %
\setlength{\tabcolsep}{6pt} %
\begin{adjustbox}{max width=\textwidth} %
\arrayrulecolor[HTML]{000000} %
\begin{tabular}{|>{\columncolor[HTML]{EFEFEF}}p{2.4cm}|p{2cm}|p{1.5cm}|p{3.7cm}|p{3.2cm}|p{2.6cm}|p{2.5cm}|}
\hline
\rowcolor[HTML]{EFEFEF} %
\multicolumn{1}{|>{\columncolor[HTML]{EFEFEF}}p{2.4cm}|}{\scriptsize \textbf{Application Area}} & 
\multicolumn{3}{c|}{\scriptsize \textbf{Annotation Type}} & 
\multicolumn{3}{c|}{\scriptsize \textbf{Annotation Generation Method}} \\ \cline{2-7}
\scriptsize \textbf{} & 
\scriptsize \textbf{Textual} & 
\scriptsize \textbf{Graphical} & 
\scriptsize \textbf{Textual and Graphical} & 
\scriptsize \textbf{Manual} & 
\scriptsize \textbf{Semi-automatic} & 
\scriptsize \textbf{Automatic} \\ \hline

\scriptsize Storytelling and Narrative Visualization & 
\scriptsize{\cite{fulda2015timelinecurator, heer2008graphical, hullman2013deeper, ko2024natural, li2023geocamera, li2023networknarratives, liew2022using, shi2024understanding, stoiber2024visahoi, zhao2021chartstory}} & 
\scriptsize{} & 
\scriptsize{\cite{bach2018design, brehmer2019timeline, bromley2023difference, bryan2016temporal, chen2023calliope, chen2023does, choe2024enhancing, dhawka2023we, eccles2008stories, endert2012semantic, fan2022annotating, gao2014newsviews, hullman2013contextifier, ibanez2023almanac, kauer2024discursive, kim2019datatoon, kwon2014visjockey, latif2021kori, lee2013sketchstory, lu2017visual, offenwanger2023timesplines, ren2017chartaccent, viegas2007manyeyes, walker2015storyboarding, wang2018narvis, wang2020cheat, wood2012sketchy, wood2018rethinking, zhao2015data}} & 
\scriptsize{\cite{choe2024enhancing, endert2012semantic, heer2008graphical, kauer2024discursive, kim2019datatoon, kwon2014visjockey, lee2013sketchstory, li2023geocamera, offenwanger2023timesplines, walker2015storyboarding, wang2020cheat, wood2018rethinking}} & 
\scriptsize{\cite{brehmer2019timeline, choe2024enhancing, eccles2008stories, endert2012semantic, fulda2015timelinecurator, hullman2013deeper, kim2019datatoon, li2023networknarratives, lu2017visual, ren2017chartaccent, viegas2007manyeyes, wang2018narvis, wood2012sketchy, zhao2015data}} & 
\scriptsize{\cite{bromley2023difference, bryan2016temporal, chen2023calliope, chen2023does, fan2022annotating, gao2014newsviews, hullman2013contextifier, ibanez2023almanac, kauer2024discursive, ko2024natural, liew2022using, shi2024understanding, stoiber2024visahoi, zhao2021chartstory}} \\ \hline

\scriptsize Collaborative Sensemaking & 
\scriptsize{\cite{abello2006ask, biehl2007fastdash, kang2014characterizing, lin2021sanguine, willett2011commentspace, wongsuphasawat2017voyager, xu2018chart}} & 
\scriptsize{\cite{wongsuphasawat2017visualizing}} & 
\scriptsize{\cite{chen2011supporting, danyluk2020touch, de2020visual, eccles2008stories, ellis2004collaborative, endert2012semantic, han2024deixis, heer2007design, heer2007voyagers, heer2012interactive, kong2009perceptual, lee2015sketchinsight, lin2022data, mathisen2019insideinsights, schwab2020visconnect, sevastjanova2021visinreport, viegas2006communication, viegas2007manyeyes, walker2015storyboarding, walny2020pixelclipper, wood2018rethinking, wright2006sandbox, zhao2016annotation, zhao2017supporting}} & 
\scriptsize{\cite{abello2006ask, danyluk2020touch, de2020visual, ellis2004collaborative, endert2012semantic, han2024deixis, heer2007voyagers, heer2012interactive, kang2014characterizing, kong2009perceptual, lee2015sketchinsight, lin2021sanguine, lin2022data, mathisen2019insideinsights, viegas2006communication, walker2015storyboarding, walny2020pixelclipper, wongsuphasawat2017voyager, wood2018rethinking, xu2018chart}} & 
\scriptsize{\cite{biehl2007fastdash, chen2011supporting, eccles2008stories, endert2012semantic, lin2022data, schwab2020visconnect, sevastjanova2021visinreport, viegas2007manyeyes, willett2011commentspace, wright2006sandbox, zhao2017supporting}} & 
\scriptsize{\cite{abello2006ask, han2024deixis, heer2007voyagers, wongsuphasawat2017visualizing, zhao2016annotation}} \\ \hline

\scriptsize Exploratory Data Analysis & 
\scriptsize{\cite{abello2006ask, collins2018guidance, fulda2015timelinecurator, gadhave2022reusing, groth2006provenance, heer2008graphical, kadivar2009capturing, kang2014characterizing, ko2024natural, sanderson1994exploratory, shi2024understanding, shrinivasan2009connecting, willett2011commentspace, wongsuphasawat2017voyager, xu2018chart, zgraggen2014panoramicdata, zhao2012facilitating}} & 
\scriptsize{\cite{eirich2021irvine, wongsuphasawat2017visualizing, zhao2011kronominer, zhao2020chartseer}} & 
\scriptsize{\cite{bach2018design, breslav2015exploratory, bryan2016temporal, choe2024enhancing, choi2015visdock, danyluk2020touch, de2020visual, elmqvist2008datameadow, endert2012semantic, franciscani2014annotation, gadhave2021predicting, han2024deixis, henry2006matrixexplorer, hoffswell2016visual, hopkins2020visualint, kandogan2012just, kong2009perceptual, kong2012graphical, kwon2014visjockey, latif2018exploring, lee2015sketchinsight, liu2018visualizing, lu2017visual, masson2023charagraph, mathisen2019insideinsights, romat2019activeink, satyanarayan2014lyra, sevastjanova2021visinreport, shrinivasan2008supporting, stasko2007jigsaw, viegas2007manyeyes, walker2015storyboarding, walny2020pixelclipper, wang2020cheat, wood2018rethinking, wright2006sandbox, xia2014annotatedtimetree, zhao2017supporting}} & 
\scriptsize{\cite{abello2006ask, breslav2015exploratory, choe2024enhancing, choi2015visdock, collins2018guidance, danyluk2020touch, de2020visual, eirich2021irvine, endert2012semantic, gadhave2022reusing, groth2006provenance, han2024deixis, heer2008graphical, kadivar2009capturing, kang2014characterizing, kong2009perceptual, kwon2014visjockey, latif2018exploring, lee2015sketchinsight, mathisen2019insideinsights, romat2019activeink, shrinivasan2008supporting, shrinivasan2009connecting, walker2015storyboarding, walny2020pixelclipper, wang2020cheat, wongsuphasawat2017voyager, wood2018rethinking, xia2014annotatedtimetree, xu2018chart, zgraggen2014panoramicdata}} & 
\scriptsize{\cite{choe2024enhancing, elmqvist2008datameadow, endert2012semantic, franciscani2014annotation, fulda2015timelinecurator, gadhave2021predicting, henry2006matrixexplorer, kong2012graphical, lu2017visual, masson2023charagraph, satyanarayan2014lyra, sevastjanova2021visinreport, stasko2007jigsaw, viegas2007manyeyes, willett2011commentspace, wright2006sandbox, zhao2011kronominer, zhao2012facilitating, zhao2017supporting, zhao2020chartseer}} & 
\scriptsize{\cite{abello2006ask, bryan2016temporal, han2024deixis, hoffswell2016visual, hopkins2020visualint, kandogan2012just, ko2024natural, latif2018exploring, liu2018visualizing, shi2024understanding, shrinivasan2009connecting, wongsuphasawat2017visualizing}} \\ \hline

\scriptsize Provenance & 
\scriptsize{\cite{bavoil2005vistrails, collins2018guidance, fujiwara2018concise, gadhave2022reusing, groth2006provenance, heer2008graphical, kadivar2009capturing, lin2021sanguine, shrinivasan2009connecting, zgraggen2014panoramicdata}} & 
\scriptsize{\cite{eirich2021irvine}} & 
\scriptsize{\cite{gadhave2021predicting, gotz2008characterizing, gratzl2016visual, heer2012interactive, lin2022data, mathisen2019insideinsights, stitz2018knowledgepearls, walker2015storyboarding}} & 
\scriptsize{\cite{collins2018guidance, eirich2021irvine, gadhave2022reusing, groth2006provenance, heer2008graphical, heer2012interactive, kadivar2009capturing, lin2021sanguine, lin2022data, mathisen2019insideinsights, shrinivasan2009connecting, walker2015storyboarding, zgraggen2014panoramicdata}} & 
\scriptsize{\cite{fujiwara2018concise, gadhave2021predicting, gratzl2016visual, lin2022data, shrinivasan2009connecting, stitz2018knowledgepearls}} & 
\scriptsize{\cite{gratzl2016visual, shrinivasan2009connecting}} \\ \hline

\scriptsize Interaction and User Engagement & 
\scriptsize{\cite{fulda2015timelinecurator, stoiber2024visahoi, zgraggen2014panoramicdata, zhao2021chartstory}} & 
\scriptsize{\cite{ritchie2019lie, xia2018dataink}} & 
\scriptsize{\cite{choe2024enhancing, de2020visual, fan2022annotating, franciscani2014annotation, han2024deixis, heer2012interactive, hoffswell2016visual, hopkins2020visualint, kauer2024discursive, kim2019datatoon, kong2012graphical, lee2013sketchstory, lin2022data, mathisen2019insideinsights, romat2019activeink, satyanarayan2014lyra, subramonyam2018smartcues, viegas2007manyeyes, walny2020pixelclipper, wang2018narvis, wood2012sketchy}} & 
\scriptsize{\cite{choe2024enhancing, de2020visual, han2024deixis, heer2012interactive, kauer2024discursive, kim2019datatoon, lee2013sketchstory, lin2022data, mathisen2019insideinsights, romat2019activeink, walny2020pixelclipper, xia2018dataink, zgraggen2014panoramicdata}} & 
\scriptsize{\cite{choe2024enhancing, fan2022annotating, franciscani2014annotation, fulda2015timelinecurator, han2024deixis, hoffswell2016visual, hopkins2020visualint, kauer2024discursive, kim2019datatoon, kong2012graphical, lin2022data, ritchie2019lie, satyanarayan2014lyra, subramonyam2018smartcues, stoiber2024visahoi, viegas2007manyeyes, wang2018narvis, wood2012sketchy, zhao2021chartstory}} & 
\scriptsize{\cite{fan2022annotating, han2024deixis, hoffswell2016visual, hopkins2020visualint, kauer2024discursive, stoiber2024visahoi, zhao2021chartstory}} \\ \hline

\end{tabular}
\end{adjustbox}
\label{table:application-annotation}
\end{table*}

Several tools have been developed to enhance provenance data analysis through the use of annotations. For instance, one model documents each step of exploration with textual and graphical annotations linked to the visualization's history~\cite{groth2006provenance}. InsideInsights integrates annotations within a collaborative environment, linking insights to specific analysis states and capturing the entire history of decision-making and iterations, enhancing transparency and reproducibility~\cite{mathisen2019insideinsights}. Similarly, a framework emphasizes semi-automated and user-guided textual annotations to document and communicate analysis steps, supporting data-driven decision-making~\cite{collins2018guidance}. A related approach uses a graphical history tool for Tableau, incorporating textual annotations to manage and recall key decision points, which enhances revisitation and sharing of the analysis journey~\cite{heer2008graphical}. Another approach enriches visual stories with contextual annotations such as text, arrows, and boxes, adaptable to different layouts~\cite{gratzl2016visual}. The Irvine system employs textual annotations to document expert knowledge while analyzing acoustic data from electrical engines, enhancing the ability to track, recall, and communicate critical insights~\cite{eirich2021irvine}.

Annotations further support provenance by adding metadata and aiding complex analyses. Some methods infer analysts' intents through annotations linked to provenance steps~\cite{gadhave2022reusing, gadhave2021predicting}. Algorithms that use user-generated annotations as keyframes help preserve critical insights and decisions while minimizing unnecessary steps, creating concise summaries of interactive network analysis~\cite{fujiwara2018concise}. Another framework leverages textual annotations for interactive network analysis, allowing users to mark key steps, which captures essential changes during exploratory analysis~\cite{fujiwara2018concise}. Additionally, tools employ textual annotations to link descriptions with visual data, improving communication in specific applications such as blood transfusion practices~\cite{lin2021sanguine}.

In summary, the integration of annotations in provenance visualization supports the documentation and communication of analytical steps, linking visual data with the analysis process. Annotations capture insights and decision points, helping to create an organized record of the workflow. However, their impact depends on how intuitively they are integrated, as excessive or complex annotations can introduce cognitive challenges.

\subsection{Interaction and User Engagement} 
Annotation techniques improve interactivity and user engagement in visualizations by allowing users to directly interact with data. Tools such as Tableau, Google Data Studio, and Microsoft Power BI enable users to add textual and graphical annotations interactively within dashboards, enhancing data exploration \cite{tableau, google_data_studio, microsoft_power_bi}. Some tools such as Click2Annotate, Touch2Annotate, and ChartAccent promote interactivity and user engagement by providing predefined or customizable annotation templates, reducing manual effort and allowing users to directly interact with visualizations, making the annotation process quicker and more accessible \cite{chen2010click2annotate, chen2010touch2annotate, ren2017chartaccent}. VisAhoi supports engagement by semi-automatically generating context-sensitive annotations using high-level visualization grammars, providing direct guidance during user interaction \cite{stoiber2024visahoi}.

The emphasis on interactive and accessible annotation processes is evident in tools such as Lyra and DataInk, which allow users to work directly on visualizations without coding, making the experience more intuitive \cite{satyanarayan2014lyra, xia2018dataink}. PanoramicData supports exploratory analysis by allowing users to manually annotate visualizations using a hybrid pen and touch system on a large 2D canvas\cite{zgraggen2014panoramicdata}. Similarly, other tools integrate digital pen and touch capabilities, enabling users to add annotations, apply filters, and highlight elements, allowing direct interaction with visual components and enhancing user engagement \cite{romat2019activeink, lee2015sketchinsight, offenwanger2023timesplines, wood2018rethinking}.

The use of collaborative and context-sensitive annotations is reflected in tools such as SmartCues and PixelClipper, where interactive overlays adjust to user actions, making annotations more relevant \cite{subramonyam2018smartcues, walny2020pixelclipper}. Viscussion links comments to specific areas of visualizations, improving context and clarity by guiding user attention without cluttering the display \cite{kauer2024discursive}. Collaborative annotation systems, such as those studied by Han et al., capture verbal and non-verbal cues during data meetings, preserving context and aiding the review process \cite{han2024deixis}. Other visualization tools that support synchronous collaboration facilitate annotations, enhancing user engagement by helping users externalize insights and actively involving them in the exploration and discussion of data~\cite{eccles2008stories, wongsuphasawat2017voyager, biehl2007fastdash}. However, it is essential to balance annotation complexity, as overly simplistic annotations can appear patronizing and may decrease user trust and engagement \cite{alebri2023embellishments}.

In summary, annotation techniques in visualizations enhance user engagement by enabling direct interaction with data, turning static visuals into dynamic and personalized experiences. These methods encourage users to explore and connect with content more deeply, while collaborative and context-sensitive annotations support engagement by adapting to user input and aiding clear communication. However, striking the right balance is essential—too much automation can potentially limit user agency, making the experience feel constrained, while excessive manual input may be cumbersome and reduce the overall usability. The focus should be on designing annotations that are intuitive and supportive, keeping users involved without overcomplicating the interaction or diminishing their control, thus maintaining an effective and engaging visualization environment.

\subsection{Uncertainty Visualizations}
Annotations contribute to the understanding and representing uncertainty in visualizations and support decision-making. Procedurally generated annotations, including variations in intensity width, exponential sharpness, and noise, help viewers comprehend uncertainty in various contexts~\cite{cedilnik2000procedural}. In meteorological applications, graphical annotations, such as color-coded lines, circles, and textual notes providing temporal and intensity data, convey information about tropical cyclone predictions, including storm size and intensity, aiding in the exploratory analysis of complex data~\cite{liu2016uncertainty, liu2018visualizing}. Furthermore, graphical annotations, including highlighted comparisons and marked extremes, support decision-making within uncertain visualizations, building user confidence with limited impact on accuracy~\cite{ferreira2014sample}.

Traditional annotation techniques for visualizing uncertainty, such as error bars, glyphs, scale modulation, and ambiguation, work well for simpler data but can lead to over-plotting in more complex visualizations. To address these challenges, annotations integrated into density and probabilistic plots offer scalable approaches that emphasize reliable data and reduce the prominence of uncertain values~\cite{feng2010matching}.

In summary, annotations are used to represent uncertainty in visualizations, supporting clarity and decision-making. Procedural methods, such as intensity variations, help convey uncertainty but may not always improve accuracy. While traditional techniques can clutter complex visuals, integrated approaches such as density and probabilistic plots focus on reliable data, reducing the prominence of uncertain values.

\setstretch{0.95}
\section{Discussion, Challenges and Future Research Directions}
\subsection{Discussion}
\label{sec.discussion}
The empirical studies (\autoref{sec.empirical-studies}) and practical applications (\autoref{sec.application}) of annotations reveal insights about their impact on comprehension, engagement, collaboration, and storytelling. Annotations provide significant benefits, but their effectiveness is tied to the context, design choices, and objectives of the visualization. We present an analysis of the general takeaways and insights, emphasizing the implications for future tool development.

\textbf{Adaptation to Context-Specific Needs:} Annotations serve multiple functions across various visualization contexts--enhancing narrative flow, aiding comprehension, and supporting collaborative analysis~\cite{kosara2013storytelling, segel2010narrative, heer2007design}. The versatility seen in tools from automated narrative systems~\cite{hullman2013contextifier, ibanez2023almanac} to highly customizable developer libraries~\cite{lu_d3-annotation, swoopyDrag} underscores the need for annotation strategies that adapt to the specific context of use. This suggests a strategic focus on aligning annotations with the purpose of the visualization and the needs of its audience rather than a one-size-fits-all approach.

\textbf{Managing Clarity and Information Overload:} Balancing the informative power of annotations with the risk of clutter is a recurring challenge. Empirical studies emphasize that annotations must be strategically placed and carefully moderated to avoid overwhelming users and to improve comprehension~\cite{stokes2022striking}. Research highlights that when positioned thoughtfully, clear and concise annotations can enhance visualizations' interpretive accuracy without detracting from the main data~\cite{ajani2021declutter, borkin2015beyond}. This emphasizes the importance of optimizing annotation density and placement, ensuring they contribute meaningfully to the visualization while maintaining clarity.

\textbf{Role of Interactivity in User Engagement:} The integration of interactive annotations in tools highlights their role in enhancing user engagement by allowing individuals to explore data actively~\cite{ren2017chartaccent, lee2013sketchstory}. This points to the value of interactivity not just as a feature but as a fundamental design principle that enables users to manipulate, adjust, and connect with the data. Emphasizing adaptable and responsive annotation features can make visualizations more engaging and informative~\cite{kauer2024discursive}.

\textbf{Preventing Misinterpretations through Design:} Annotations have the potential to clarify or confuse, depending on their design. Findings stress that poor design can mislead viewers, especially in areas where data misinterpretation carries substantial consequences~\cite{lisnic2023misleading, ge2023calvi}. This reinforces the need for responsible design practices, ensuring annotations are precise, transparent, and contextually relevant to prevent miscommunication and preserve the integrity of the visualization.

\textbf{Automation as a Double-Edged Tool:} Automation offers consistency and efficiency in generating annotations but requires careful implementation to ensure relevance and contextual accuracy. The analysis of automated systems~\cite{bryan2016temporal, chen2023calliope} shows that they can streamline annotation processes; however, they must be balanced with user input to refine and validate outputs. This suggests a hybrid approach that combines machine efficiency with human oversight for optimal results.

\subsection{Challenges and Future Research Directions}

The current state of annotation research and applications reveals both strengths and gaps in our understanding of how annotations enhance visualizations. Empirical studies offer insights into the various benefits of annotations (see \autoref{sec.empirical-studies}). However, further research is needed to address gaps that will help develop better annotation tools, making visualizations more effective, intuitive, and adaptable to diverse user needs.

\vspace{-3pt}
\noindent
\begin{center}
\textbf{Optimal Placement and Quantity of Annotations \\ for Effective Visualization Comprehension}
\end{center}
\vspace{-3pt}

\textit{Literature Gap \& Challenge:} As noted in \autoref{sec.discussion}, balancing the informative power of annotations with the risk of visual clutter is a persistent challenge, with studies emphasizing strategic placement to avoid overwhelming users~\cite{stokes2022striking, ajani2021declutter}. However, there is limited research specifically focused on identifying the optimal placement and quantity of annotations across different contexts. Cartography provides established practices for label placement~\cite{ooms2012investigating, kern2008automation}, but similar strategies tailored to data visualizations remain underexplored, especially regarding the integration of both textual and graphical elements~\cite{kosara2013storytelling, ottley2019curious}.

\textit{Research Direction:} Future research should investigate how variations in annotation placement and density impact user comprehension using methods such as eye-tracking and controlled user studies. Developing adaptable guidelines tailored to visualization types and audience needs will directly address the issues of clarity and overload discussed earlier, helping to refine annotation strategies for enhanced visual communication.

\vspace{-3pt}
\noindent
\begin{center}
\textbf{How Professionals Use Annotations in Visualization}    
\end{center}
\vspace{-3pt}

\textit{Literature Gap \& Challenge:} The visualization community has long sought to understand how practitioners design and use visualizations in real-world scenarios~\cite{wang2020cheat, parsons2021understanding, parsons2020data, li2023knowledge}. Although many tools and studies draw inspiration from practitioners, significantly influencing annotation research in visualizations~\cite{ren2017chartaccent, hullman2013contextifier, segel2010narrative, hullman2011visualization, rahman2023exploring}, a gap remains in understanding the practical aspects of annotation use. We lack direct input from professionals
who use annotations to communicate insights to broader audiences. This disconnect limits the integration of real-world practices into theoretical frameworks and tools supporting annotations.

\textit{Research Direction:} To bridge this gap, qualitative research involving interviews and case studies with visualization professionals, including data journalists, data scientists, and data analysts, is essential. The focus of this research should be to understand their specific methods, preferences, and the challenges they encounter when annotating visualizations. The insights derived from such studies have the potential to guide the development of new tools or methodologies that address the identified challenges, connecting theoretical knowledge with practical application in the field of data visualization.

\vspace{-3pt}
\noindent
\begin{center}
\textbf{Understanding the Impact of Annotations on Comprehension in Varying Context}
\end{center}
\vspace{-3pt}

\textit{Literature Gap \& Challenge:} In \autoref{sec.discussion}, we highlight the importance of context-specific annotation strategies and clarity management to avoid information overload. Although empirical studies show that annotations enhance comprehension, memorability, and reduce cognitive load~\cite{borkin2015beyond, ottley2019curious}, their effectiveness may vary with visualization type, task complexity, domain knowledge, and users' visualization literacy~\cite{amarstasko, guo2024we, chun2020giving}. Designing annotations that address these diverse factors remains challenging. Additionally, Quadri et al. identified a disconnect between designers' communication goals and audience interpretation~\cite{quadri2024you}, emphasizing the need for further study to explore how annotations impact comprehension across varied contexts.

\textit{Research Direction:} Future investigations should look at how different annotation styles, levels of detail, and placement strategies affect comprehension across various visualization contexts. Studies should include participants with diverse visual literacy levels and domain expertise to understand how these factors influence interpretation and decision-making. Approaches such as interviews, think-aloud, and case studies can offer insights into user interactions, whereas larger empirical studies can identify trends in comprehension and usability. The aim is to develop evidence-based guidelines that tailor annotation strategies to specific visualization types and user needs, enhancing clarity and accessibility.

\vspace{-3pt}
\noindent
\begin{center}
\textbf{Automatic Annotation Generation for \\ Better Tooling Support}
\end{center}
\vspace{-3pt}

\textit{Literature Gap \& Challenge:} In \autoref{sec.discussion}, we emphasize automation’s role in enhancing annotation consistency and efficiency. The visualization community has advanced automatic annotation generation using linguistic analysis, statistical methods, signal processing, and semantic models~\cite{bromley2023difference, ibanez2023almanac, kandogan2012just, hullman2013contextifier}. With the rise of LLMs for automated content generation, researchers are exploring their use in visualization~\cite{choe2024enhancing, xu2024exploring, ye2024generative, ko2024natural, liew2022using}. Expanding on these efforts, LLMs can be used to generate context-aware, personalized annotations, improving tool support for annotation authoring.

\textit{Research Direction:} Future research should focus on integrating LLMs with visualization tools to automate annotation. LLMs have the potential to interpret and describe visual data, generating context-aware annotations that could reduce manual work and simplify complex information. Future research should make efforts to evaluate and use LLMs to improve the automatic generation of annotations while further refining traditional methods like rule-based systems and machine-learning models.

\vspace{-3pt}
\noindent
\begin{center}
\textbf{Establishing Universal Principles for an \\ Annotation Design Space}
\end{center}
\vspace{-3pt}

\textit{Literature Gap \& Challenge:}
While we emphasize the importance of context-specific annotation strategies in \autoref{sec.discussion}, there remains a need for common principles that can be adapted across various visualization scenarios. Existing research often addresses annotations in limited contexts. For instance, Ren et al. focused on specific chart types such as bar charts, line charts, and scatterplots, and Hullman et al. explored textual annotations without considering graphical elements~\cite{ren2017chartaccent, hullman2013contextifier}. Rahman et al. examined multiple chart types but restricted their analysis to static visualizations, overlooking dynamic contexts where interactivity plays a significant role~\cite{rahman2024qualitative}. To create a unified design framework, it is essential to incorporate insights from the empirical studies discussed earlier, such as the impact of strategic annotation placement, density, and style on avoiding clutter and enhancing comprehension across different user groups. By synthesizing these findings, we can establish adaptable guidelines that balance context-specific needs with broadly applicable principles, ensuring effective use across diverse visualization scenarios.

\textit{Research Direction:} Future research should work towards creating a design space for annotations that consolidates universal guidelines applicable across various visualization contexts. This design space should draw from empirical studies on visual literacy, accessibility, and dynamic visualizations, offering principles that balance cognitive load, user comprehension, and adaptability. The goal is to develop a set of foundational guidelines that can be flexibly applied, supporting the creation of effective annotation strategies tailored to specific needs while maintaining coherence and usability in diverse visualization scenarios.

\section{Conclusion}
\label{sec.conclusion}
We conducted a comprehensive survey on annotations in information visualization, exploring their types, design spaces, and generation methods across various tools and techniques facilitating annotations. Our study synthesizes empirical findings on annotations, identifies their practical applications, and examines the implications of existing research to improve tool development. We discuss existing challenges and propose future research directions aimed at improving annotation practices, with the goal of enabling the visualization community to create more effective data communication tools.

\section*{Acknowledgment}
\addcontentsline{toc}{section}{Acknowledgment}

We used the generative AI tool ChatGPT 3.5 by OpenAI to edit, paraphrase, and restructure the text, which was later reviewed and revised. Figures from outside sources have been cited and used in accordance with the fair use doctrine. This work was partially supported by NSF IIS-2316496 and DUE-2216227.

\setstretch{0.95}
\bibliographystyle{IEEEtran}

\setstretch{0.95}

\section{Biography Section}
\label{sec.bibliography}

\begin{IEEEbiography}[{\includegraphics[width=1in,height=1.25in,clip,keepaspectratio]{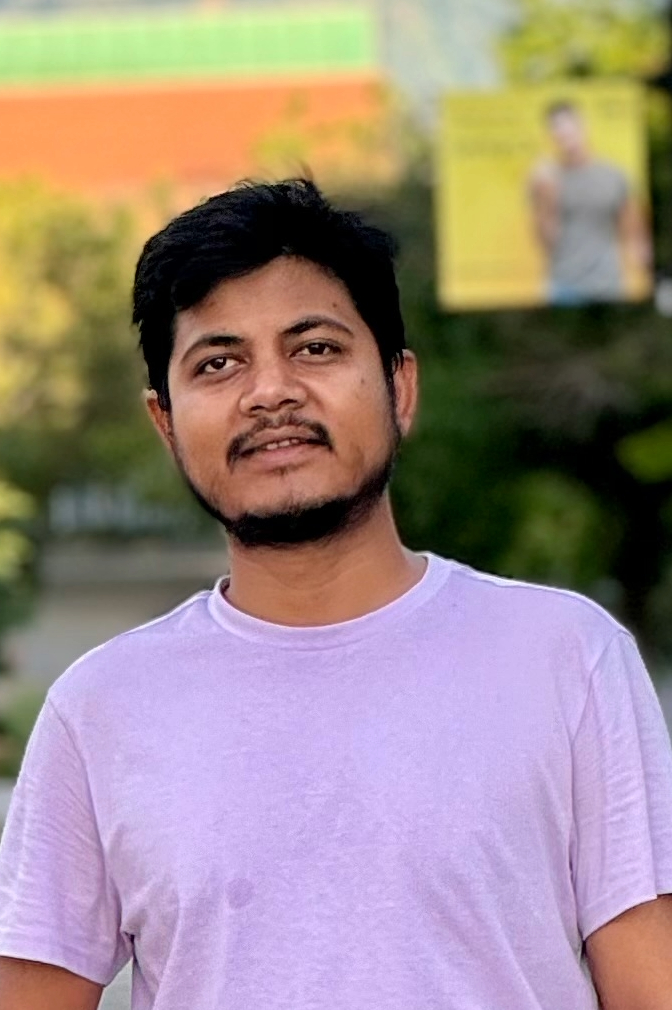}}]{Md Dilshadur Rahman} is pursuing a PhD in the Kahlert School of Computing and the SCI Institute at the University of Utah. Prior to joining the University of Utah, he completed an MS in Computer Science from the University of South Florida. Before his graduate studies, he served as a Computer Science lecturer at Daffodil International University in Bangladesh. His research focuses on Information Visualization and Human-Computer Interaction.
\end{IEEEbiography}

\begin{IEEEbiography}[{\includegraphics[width=1in,height=1.25in,clip,keepaspectratio]{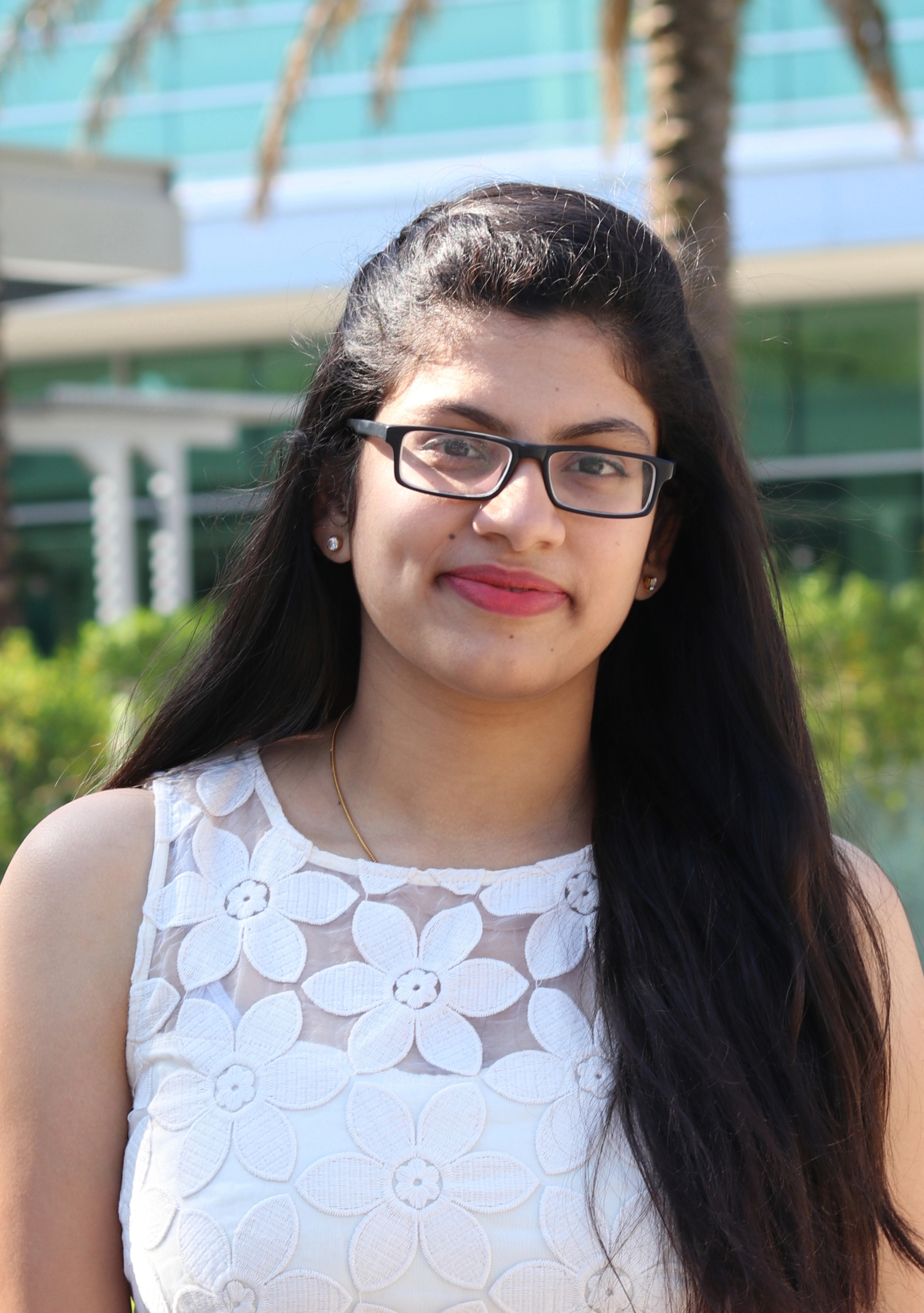}}]{Bhavana Doppalapudi} is a PhD student at the Department of Computer Science and Engineering at the University of South Florida. Prior to joining as a PhD student at the University of South Florida, Bhavana completed her Master from the same institution. Her research is focused on the impact of visualizations in Decision Making in Machine Learning.
\end{IEEEbiography}

\begin{IEEEbiography}[{\includegraphics[width=1in,height=1.25in,clip,keepaspectratio]{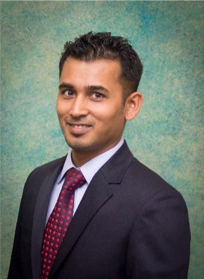}}]{Ghulam Jilani Quadri} is an Assistant Professor in the School of Computer Science at the University of Oklahoma. Prior to that, Dr. Quadri was a CIFellow postdoc at the University of North Carolina at Chapel Hill. He received his Ph.D. from the University of South Florida. Quadri received the 2021 Computing Innovation Fellow award. His research interests include creating human-centered frameworks to optimize visualization design and improve decision-making quality.
\end{IEEEbiography}

\begin{IEEEbiography}[{\includegraphics[width=1in,height=1.25in,clip,keepaspectratio]{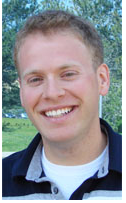}}]{Paul Rosen} is an Associate Professor at the University of Utah. He received his Ph.D.\ from Purdue University. His research interests include applying geometry- and topology-based approaches to problems in information visualization. Along with his collaborators, he has received best paper awards or honorable mentions at IEEE VIS, IEEE PacificVis, CG\&A, IVAPP, and SIBGRAPI.  Dr.\ Rosen received a National Science Foundation CAREER Award in 2019.
\end{IEEEbiography}

\vfill

\vfill

\end{document}